\documentclass[conference,a4paper]{IEEEtran}
\IEEEoverridecommandlockouts
\usepackage{url}
\usepackage{cite}
\usepackage{amsmath,amssymb,amsfonts}
\usepackage{algorithmic}
\usepackage{graphicx}
\usepackage{textcomp}
\usepackage{xcolor}
\usepackage{etoolbox}
\usepackage{dblfloatfix}
\usepackage{hyperref}
\def\BibTeX{{\rm B\kern-.05em{\sc i\kern-.025em b}\kern-.08em
    T\kern-.1667em\lower.7ex\hbox{E}\kern-.125emX}}
    
\newcommand\ChangeRT[1]{\noalign{\hrule height #1}}

\begin{document}

\bstctlcite{IEEEexample:BSTcontrol}

\title{TW-BAG: Tensor-wise Brain-aware Gate Network for Inpainting Disrupted Diffusion Tensor Imaging}

\author{\IEEEauthorblockN{
Zihao Tang\IEEEauthorrefmark{1}\IEEEauthorrefmark{2},
Xinyi Wang\IEEEauthorrefmark{1}\IEEEauthorrefmark{2},
Lihaowen Zhu\IEEEauthorrefmark{1}, 
Mariano Cabezas\IEEEauthorrefmark{2},
Dongnan Liu\IEEEauthorrefmark{1}\IEEEauthorrefmark{2},
Michael Barnett\IEEEauthorrefmark{2}\IEEEauthorrefmark{3},\\
Weidong Cai\IEEEauthorrefmark{1},
and Chenyu Wang\IEEEauthorrefmark{2}\IEEEauthorrefmark{3}}

        \IEEEauthorblockA{\IEEEauthorrefmark{1}School of Computer Science, University of Sydney, Australia}
        \IEEEauthorblockA{\href{mailto:zihao.tang@sydney.edu.au}{zihao.tang@sydney.edu.au}}
        \IEEEauthorblockA{\IEEEauthorrefmark{2}Brain and Mind Centre, University of Sydney, Australia}
        \IEEEauthorblockA{\IEEEauthorrefmark{3}Sydney Neuroimaging Analysis Centre, Australia}
}

\maketitle

\begin{abstract}
Diffusion Weighted Imaging (DWI) is an advanced imaging technique commonly used in neuroscience and neurological clinical research through a Diffusion Tensor Imaging (DTI) model. Volumetric scalar metrics including fractional anisotropy, mean diffusivity, and axial diffusivity can be derived from the DTI model to summarise water diffusivity and other quantitative microstructural information for clinical studies. However,  clinical practice constraints can lead to sub-optimal DWI acquisitions with missing slices (either due to a limited field of view or the acquisition of disrupted slices). To avoid discarding valuable subjects for group-wise studies, we propose a novel 3D Tensor-Wise Brain-Aware Gate network (TW-BAG) for inpainting disrupted DTIs. The proposed method is tailored to the problem with a dynamic gate mechanism and independent tensor-wise decoders. We evaluated the proposed method on the publicly available Human Connectome Project (HCP) dataset using common image similarity metrics derived from the predicted tensors and scalar DTI metrics. Our experimental results show that the proposed approach can reconstruct the original brain DTI volume and recover relevant clinical imaging information.
\end{abstract}


\section{Introduction}
\begin{figure}[t]
\centering
\begin{tabular}{cc}
\includegraphics[width=0.1\textwidth]{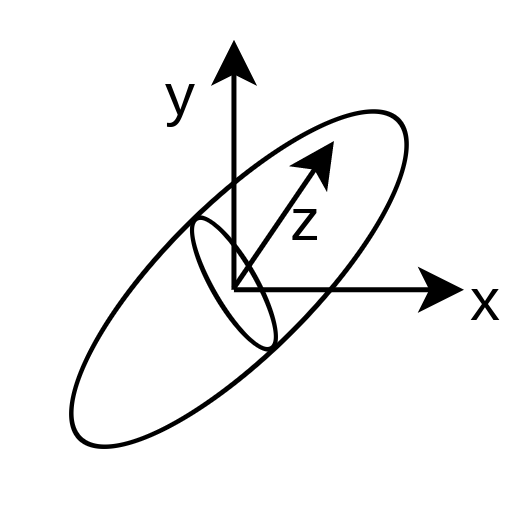} &
\includegraphics[width=0.25\textwidth]{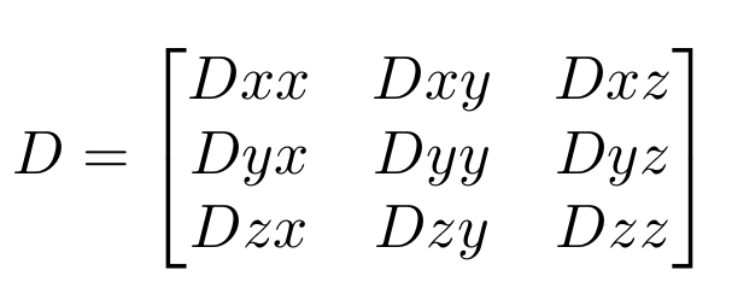} \\
\small (a) Diffusion ellipsoid & \small (b) Diffusion tensor \\[6pt]
\end{tabular}

\caption{Diffusion representations in ellipsoid and tensor.}
\label{fig:tensor_explain}
\end{figure}

Diffusion Weighted Imaging (DWI) is a non-invasive magnetic resonance imaging (MRI) modality developed to monitor water diffusivity and reveal the micro-structure of the human body~\cite{baliyan2016diffusion}. While conventional structural MRI sequences (e.g. T1, T2) provide information about the general morphological information of the brain, DWI provides additional information regarding brain dynamics with water diffusivity as a proxy. This set of sequences acquired with different gradient directions can then be summarised using a Diffusion Tensor Imaging (DTI) model to uncover microstructural information. Based on the fact that different tissues have different diffusion properties, the DTI model explains the directionality of the water diffusivity and its corresponding quantitative anisotropy~\cite{soares2013hitchhiker}. The diffusion of a particular voxel can be characterized as an ellipsoid (Figure~\ref{fig:tensor_explain}(a)) that can be mathematically formulated as a symmetric $3 \times 3$ tensor matrix (Figure~\ref{fig:tensor_explain}(b)). The coordinate system is typically aligned with the main magnetic field and body of the patient. Since the relationship between two principal directions is theoretically symmetric, i.e., $Dxy = Dyx$, $Dxz = Dzx$ and $Dyz = Dzy$, only 6 unique coefficients are needed to construct the tensor at a particular voxel.

To better understand the properties of the tensor, several scalar metrics can be derived to quantify its ellipsoid shape for comparison in groupwise analysis~\cite{feldman2010diffusion}.  Axial Diffusivity (AD), Mean Diffusivity (MD), and Fractional Anisotropy (FA) are the most commonly used scalar metrics in clinical settings and neurological research. FA and MD are measures of relative anisotropy and average magnitude of diffusion at a voxel, respectively. AD represents the magnitude of diffusion parallel to fibre tracts, assuming that a single fibre tract is available and is represented by the principal direction of the tensor. Examples of research on neurological disorders using these metrics include studies of multiple sclerosis (MS), amyotrophic lateral sclerosis (ALS), Alzheimer's dementia (AD), Parkinson's disease (PD), epilepsy, and other disease types resulting in brain damage~\cite{ma2022multiple,tae2018current}. In general,  FA decreases in the regions impacted by pathological factors (such as oedema, demyelination, gliosis, and inflammation~\cite{assaf2008diffusion}) that constitute discriminative biomarkers between patients and controls~\cite{sbardella2013dti}. For example, DTI has been proven to identify multiple sclerosis lesions with severe tissue damage and monitor tissue changes~\cite{filippi2001diffusion}. Specifically, the patient group had a higher MD and lower FA in the normal appearing white matter~\cite{ciccarelli2001investigation}.

\begin{figure*}[htbp]
\centerline{\includegraphics[width=\textwidth]{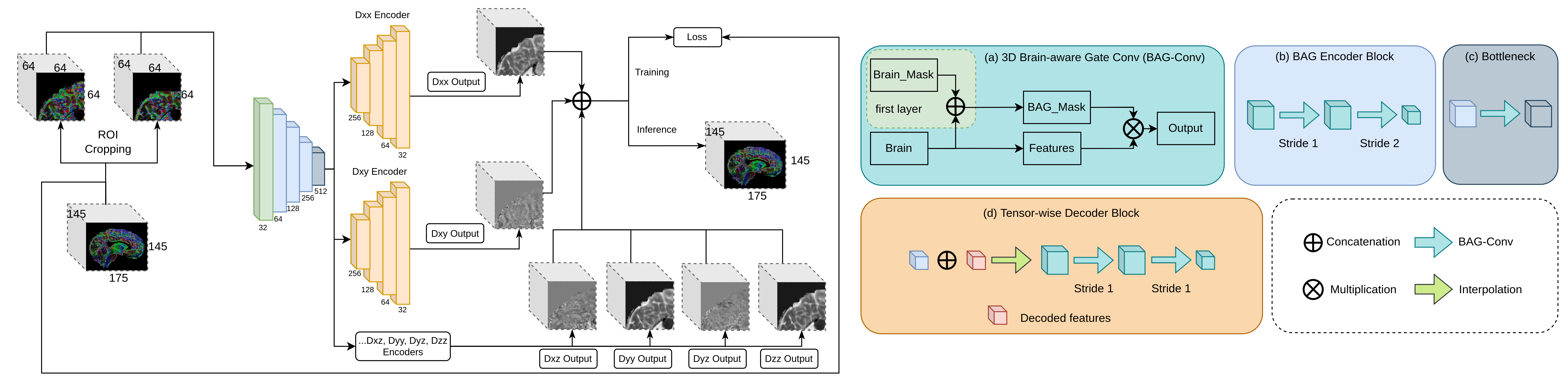}}
\caption[Pipeline]{The framework of the proposed 3D TW-BAG Network for DTI inpainting. (a) 3D Brain-aware Gate Conv (BAG-Conv) to learn a dynamic mask that guides the network to only learn the information from valid regions. (b) The architecture of the BAG encoder block. (c) The architecture of the bottleneck between the encoding and decoding stages. (d) The architecture of the decoder for each tensor model coefficient ($Dxx$, $Dxy$, $Dxz$, $Dyy$, $Dyz$, and $Dzz$).}
\label{fig:framework}
\end{figure*}

However, the acquisition can be sub-optimal due to clinical imaging constraints, such as acquisition time or patient movement during the MRI session. Unlike the structural imaging that usually requires a single acquisition, DWI requires multiple sequences with varying gradient pulses. The larger the number of gradient pulses needed in the scanning phase, the higher the angular resolution of the output DWI will have, which leads to a better representation of microstructure at the expense of a longer scanning time and higher cost. As a trade-off in clinical imaging practice, the field of view (FOV) for each subject might be reduced due to time and cost constraints, thus often resulting in missing slices from the superior part of the brain (this scenario is illustrated in Figure~\ref{fig:framework}). Another common scenario is the unwanted disruption of the images and signal caused by different types of artifacts and noise including motion, susceptibility distortions, gradient non-linearity and eddy currents. A typical pipeline for DWI usually includes correction algorithms for these perturbations, however, even those methods can sometimes fail on particular cases leading to their exclusion from the study in real world scenarios. All these factors lead to disrupted DTIs and cascading errors as a result. For example, scalar metrics will be affected at the terminal points of the frontal and parietal lobes if the top part of the brain is cropped~\cite{duffau2014dangers} and as mentioned, these subjects would fail quality control and be rejected for any further use~\cite{charlton2006white}. Hence a solution for the problems listed above is necessary to obtain consistent and spatially-aware scalar metrics from disrupted DTIs for clinical studies without discarding valuable scans. Inspired by inpainting research, the solution we propose in this work is to reframe the problem as the inpainting of the disrupted regions in DTI volumes.

The aim of image inpainting is to fill the missing areas of images with the most likely estimates based on valid information (e.g. neighboring intensity values or general image structure). Due to its generality, it can be applied to a wide range of real world applications, such as repairing old or damaged pictures~\cite{wan2020bringing} or removing specific objects~\cite{yu2019free}. Traditional approaches matched the region of interest (ROI) with the most similar patches in the remaining part of the input image~\cite{barnes2011patchmatch}. With the advancement of deep learning techniques, the overall quality of inpainted images has been improved dramatically. One of the most widely used network structures for any medical imaging task is the U-Net\cite{ron2015unet} architecture. One of its advancements was the introduction of skip connections between an encoder and a decoder to combine contextual information captured at different levels of a resolution pyramid. Due to its strengths, the U-Net has become a de-facto standard architecture of deep learning and its extensions have achieved state-of-the-art in many different areas of image analysis, including inpainting~\cite{wan2020bringing}. Another widely used technique that was proposed at the beginning of the current trend of deep learning techniques was the adversarial training strategy~\cite{goodfellow2014generative}. Generative Adversarial Network (GAN) has rapidly become one of the most popular deep learning algorithms for image generation tasks and inpainting~\cite{creswell2018generative,sal2021a}. The strategy proposes the inclusion of a discriminator to distinguish between images generated by the backbone network and the real samples. Through natural competition between the generator and discriminator networks, a Nash equilibrium can be reached where the real and predicted images can no longer be distinguished. Due to its superior performance, GAN variants and improvements were proposed based on the original to improve shortcomings of the original proposal. For example, conditional GANs introduce the use of additional input information to condition the generator and discriminator and constrain the output images~\cite{mirza2014conditional} instead of producing random types of output images. Other advancements, relevant to DTI due to its high dimensionality, are patch-based extensions to break the input image into smaller $N \times N$ patches \cite{isola2017image}.

While we have presented a short overview on natural images, image generation tasks have also been extensively used in medical imaging in recent years for specific purposes in different scenarios. For example, medical image segmentation could be reformulated as inpainting to better learn the features of ROIs~\cite{zhang2018ms,liu2019nuclei} and improve the final masks, and image synthesis could be used to enlarge the dataset due to low quality and quantity of the training data~\cite{jin2018ct}. In general, medical images can be disrupted by specific pathology~\cite{battaglini2012evaluating}, image distortions~\cite{rohde2004comprehensive}, or sub-optimal acquisition protocols~\cite{ayub2020inpainting}. Consequently, inpainting has become a popular solution to reconstruct the disrupted images to improve the accuracy of downstream clinical analysis~\cite{tang2021lg,ayub2020inpainting}.

Although GAN-based methods can achieve satisfying results on natural images, the original implementation and its variations have limitations on their training stability due to the introduction of complex auxiliary architectures and their corresponding unstable losses that can lead to mode collapse~\cite{metz2017unrolled, arjovsky2017wasserstein}. This problem becomes amplified when applied to 3D medical images due to the increased complexity and dimensionality. To this end, we propose Tensor-Wise Brain-Aware Gate (TW-BAG) Network, which is specifically designed for inpainting of missing slices in DTI images\footnote{The repository for this work is available at: \url{https://mri-synthesis.github.io}}. The overall framework of TW-BAG Network is shown in Figure~\ref{fig:framework}. To the best of our knowledge, TW-BAG is the first work that inpaints the cropped regions directly on DTI volumes, rather than the raw DWI sequence. Our experimental results show that TW-BAG is able to obtain satisfactory inpainting results in terms of computer vision metrics and from a clinical perspective. Extensive analysis has also been conducted on the efficiency of the model by feeding different percentages of valid brain training regions into the network to explore how much global brain information can help to reconstruct the missing slices and recover quantitative clinical information. 

The remainder of the paper is organized as follows. In Section~\ref{sec:methods}, we describe the proposed TW-BAG strategy in details. In Section~\ref{sec:results}, we provide the dataset used to perform the experiments, experimental settings and corresponding results including an ablation study, an efficiency study and a discussion on the clinical impact of these results. Finally, in Section~\ref{sec:conclusions}, the conclusions are presented.

\section{Methods}
\label{sec:methods}

\subsection{Problem Definition}
The data samples are 4D arrays with the first dimension representing the tensor unique values ordered as follows: $Dxx$, $Dxy$, $Dxz$, $Dyy$, $Dyz$, and $Dzz$, where $Dxx$, $Dyy$, and $Dzz$ represent the diffusion coefficients along the x, y, and z axis; and $Dxy$, $Dyz$, and $Dxz$ reflect the correlation of the random movement between each pair of directions. To evaluate the effect of our proposed inpainting method in a controlled environment, we synthesized the effect of missing slices in DTIs by setting the values in the cropped ROIs to 0 for all the subjects. The dynamic cropped range was adjusted to the first valid axial brain slice from the top of the brain and the following 10\% of the following brain slices (15 slices) in the superior-to-inferior direction of an axial plane acquisition. The proposed TW-BAG is designed to reconstruct the tensor coefficients in these cropped regions.

\subsection{Tensor-wise Brain-Aware Gate Network}
In clinical practice, high quality T1-weighted images are easier to acquire than DWI due to a much shorter scanning time and less potential distortion factors. Thus, structural images are commonly acquired by default and their availability in any diffusion study can be assumed. Therefore, brain masks can be easily generated from T1 images and then co-registered to DWI. The proposed brain-aware gate decoder follows that assumption to introduce a valid brain mask into the network during the training phase. The disrupted DTI volumes and their corresponding brain masks are concatenated before feeding them to a brain-aware gate (BAG) encoder block. Each block is composed of two parallel branches: one of them focuses on learning a dynamic mask to better represent the features, while the other follows a regular 3D convolutional path. The 3D BAG convolution is summarized as:

\begin{equation}
    BAG\_Mask_{n} = 
    \begin{cases}
    n=1,   Conv_{mask}(Brain \oplus Brain\_Mask )\\
    n>1,   Conv_{mask}(O_{n-1})\\
    \end{cases}
\end{equation}
and $O_{n}$ is defined as:
\begin{equation}
    O_{n} = \delta(Conv_{feat}(O_{n-1})) \odot \sigma (BAG\_Mask_{n}),
\end{equation}

\noindent where $\delta$ can be any activation function (LeakyReLU in this work) and $\sigma$ is the sigmoid function that provides a soft mask. Each BAG encoder consists of two consecutive BAG convolutions with stride and dilation set to 1. The BAG encoder consists of 4 BAG encoder blocks with output channels of 32, 64, 128, 256 and one bottleneck block with output channel of 512 as illustrated in Figure~\ref{fig:framework}.

\begin{figure*}[t]
\centering
\begin{tabular}{ccccccc}
\includegraphics[width=0.12\textwidth]{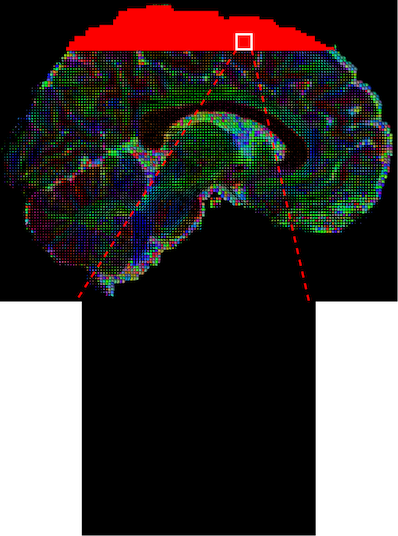} &
\includegraphics[width=0.12\textwidth]{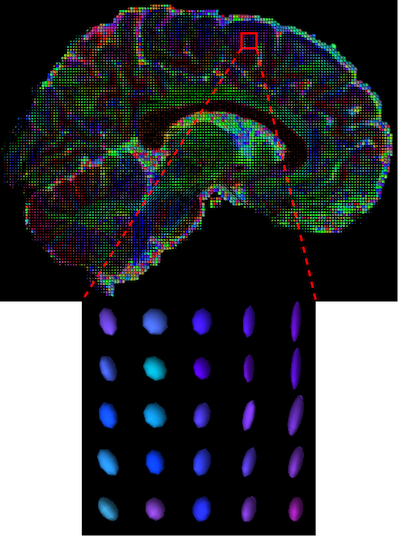} &
\includegraphics[width=0.12\textwidth]{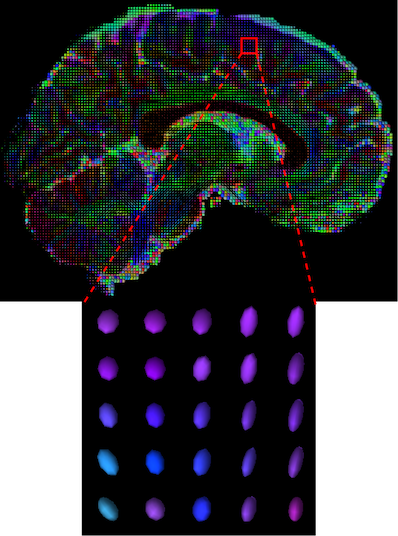} &
\includegraphics[width=0.12\textwidth]{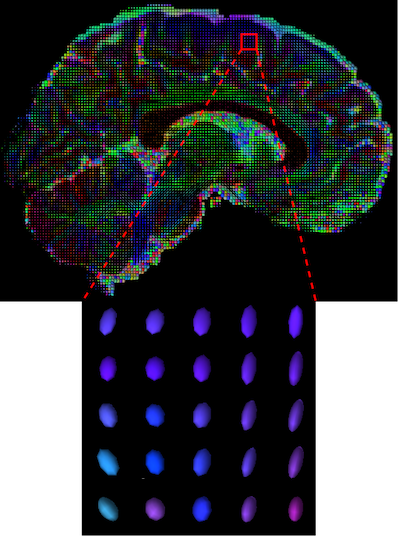} &
\includegraphics[width=0.12\textwidth]{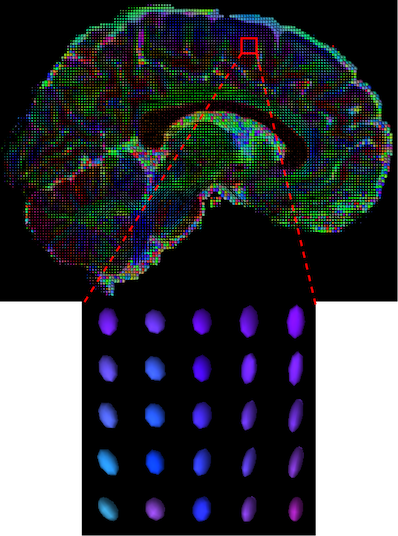} &
\includegraphics[width=0.12\textwidth]{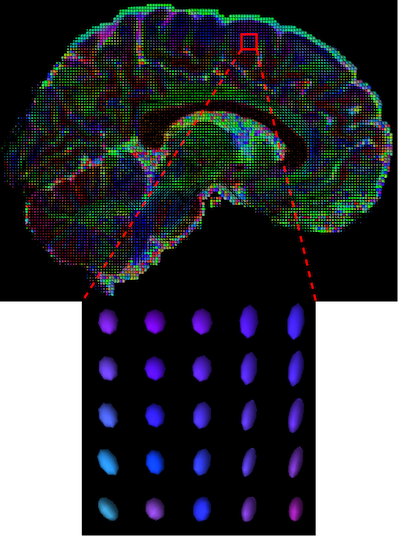} &
\includegraphics[width=0.12\textwidth]{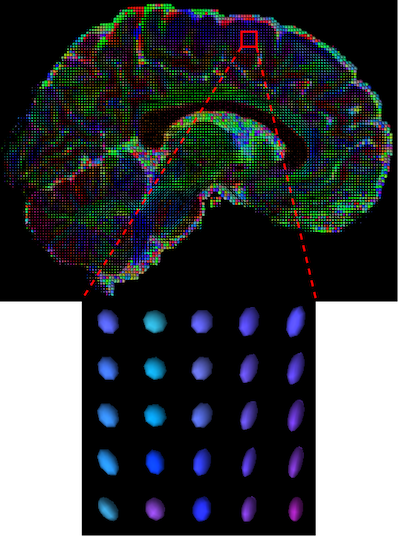}\\
\end{tabular}
\begin{tabular}{ccccccc}
\includegraphics[width=0.12\textwidth]{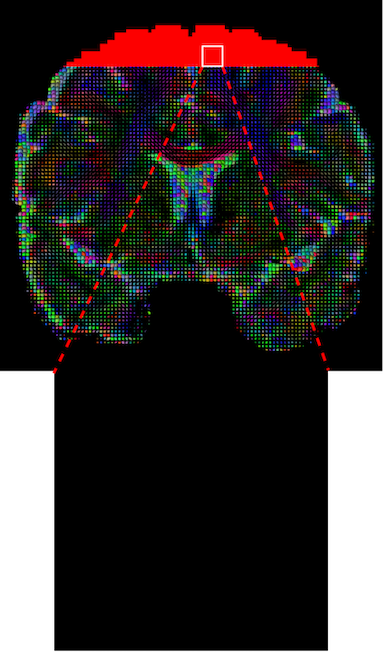} &
\includegraphics[width=0.12\textwidth]{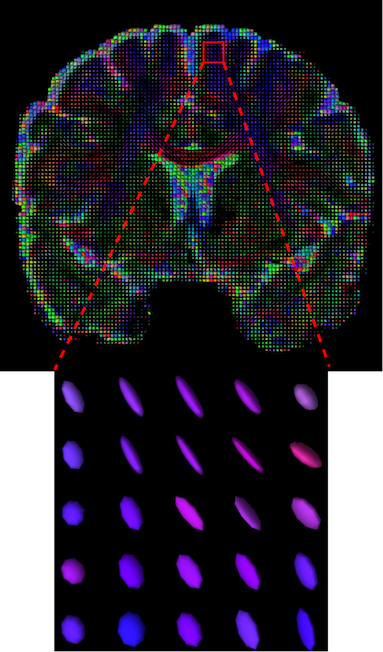} &
\includegraphics[width=0.12\textwidth]{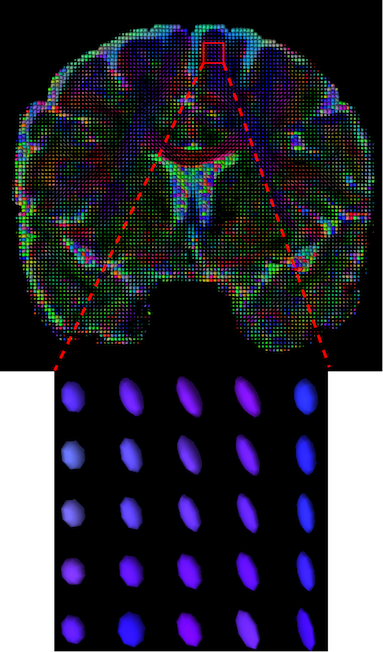} &
\includegraphics[width=0.12\textwidth]{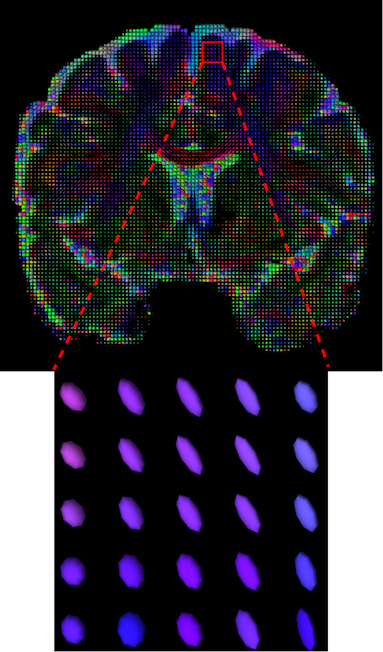} &
\includegraphics[width=0.12\textwidth]{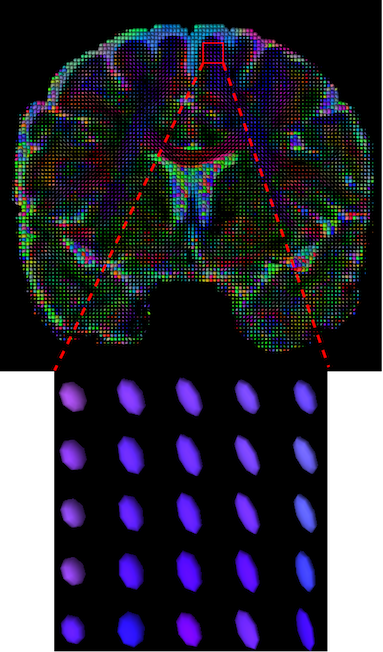} &
\includegraphics[width=0.12\textwidth]{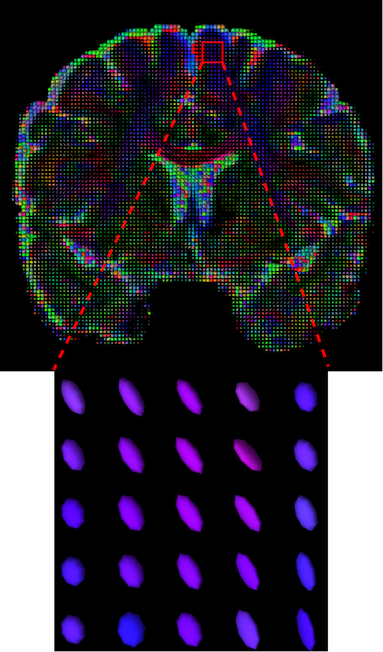} &
\includegraphics[width=0.12\textwidth]{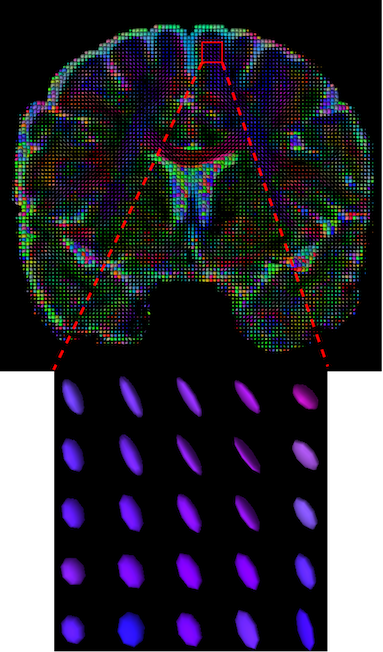}\\
\end{tabular}
\begin{tabular}{ccccccc}
\includegraphics[width=0.12\textwidth]{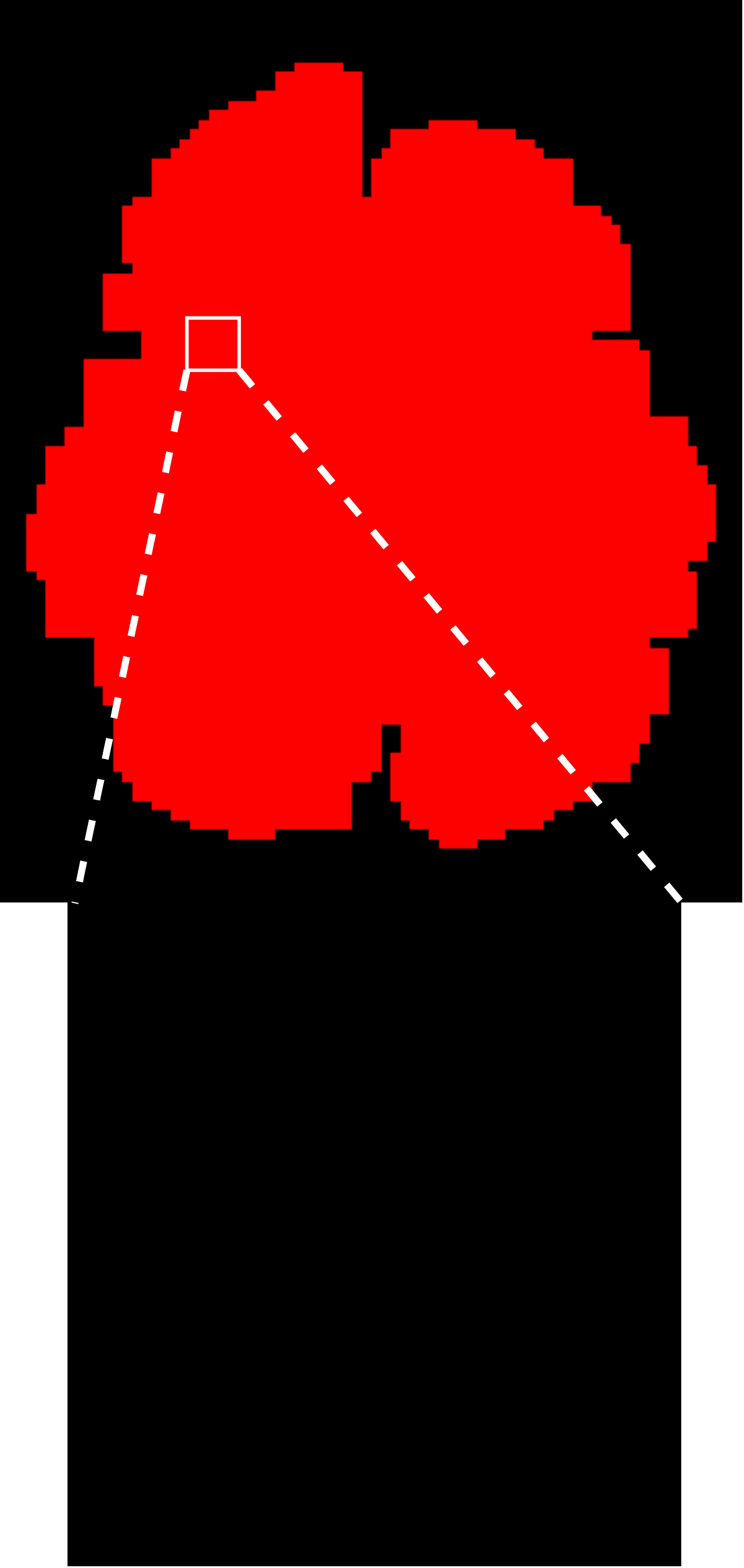} &
\includegraphics[width=0.12\textwidth]{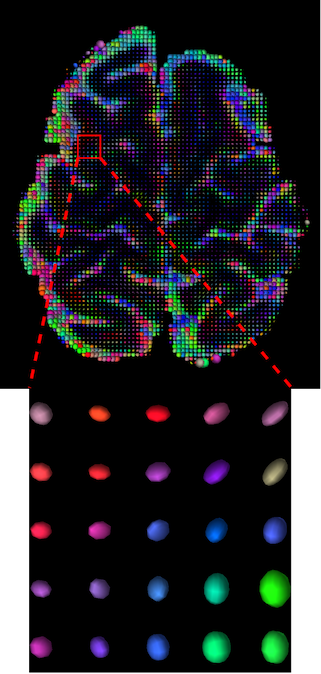} &
\includegraphics[width=0.12\textwidth]{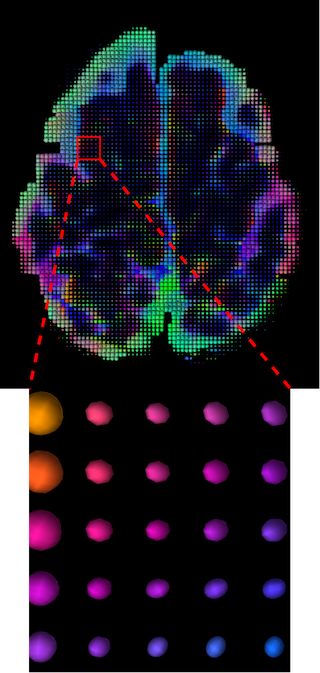} &
\includegraphics[width=0.12\textwidth]{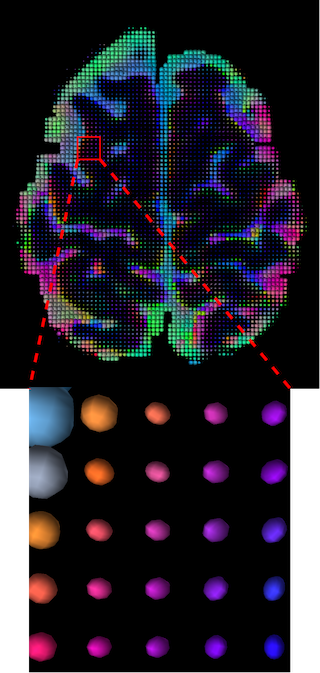} &
\includegraphics[width=0.12\textwidth]{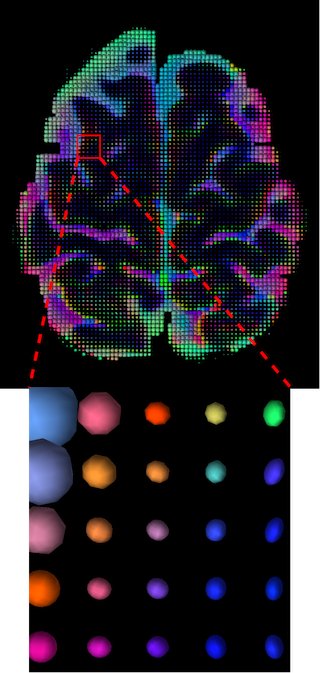} &
\includegraphics[width=0.12\textwidth]{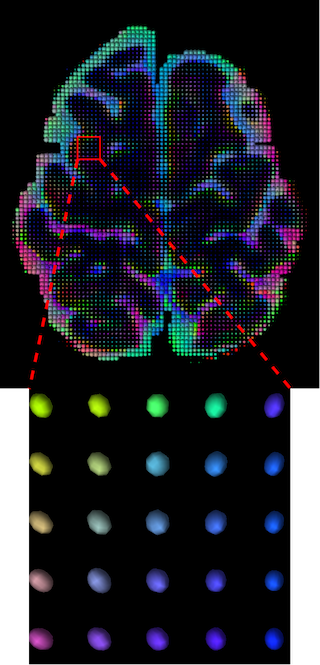} &
\includegraphics[width=0.12\textwidth]{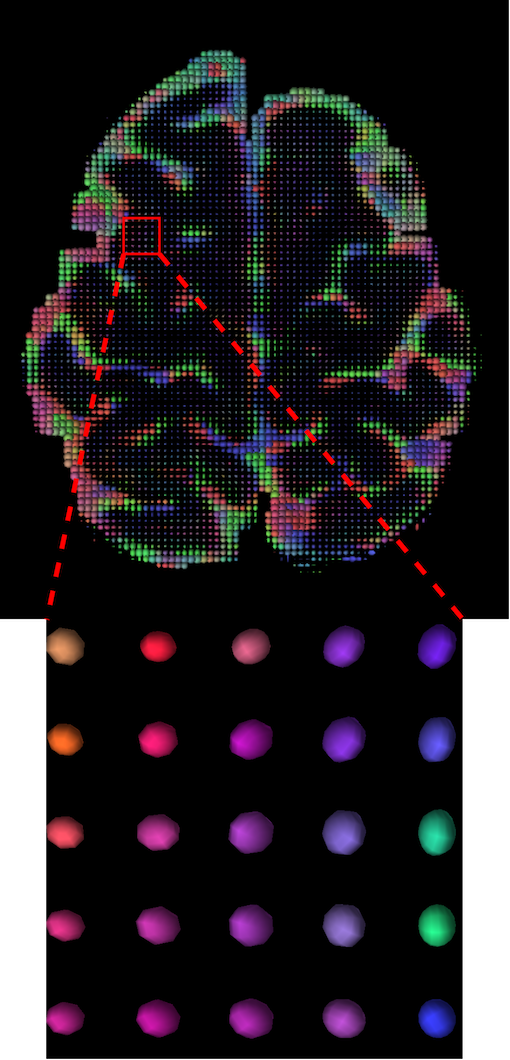}\\
\small (a) Disrupted & \small (b) GT & \small (c) U-VQVAE & \small (d) Baseline & \small (e) w/o BA-TW & \small (f) w/o TW & \textbf{(g) TW-BAG} \\[6pt]
\end{tabular}
\caption{Inpainted DTI results obtained by the compared methods with (a) as the input of the network with the values in the disrupted region (marked in red) set to 0. (b) is the ground truth image, (c)-(g) are the corresponding inpainted DTIs generated by the U-VQVAE, our proposed method without BAG-TW, BA-TW, TW and the proposed TW-BAG, respectively.}
\label{fig:vis_tensor}
\end{figure*}

As mentioned in the introduction, the diffusion tensor model can be characterized by six unique diffusion coefficients per voxel ($Dxx$, $Dxy$, $Dxz$, $Dyy$, $Dyz$, and $Dzz$). By definition, each diffusion coefficient is independent from the others. Thus the value ranges of tensor coefficients are different for each other. To this end, we propose to use independent tensor-wise (TW) coefficient specific decoders. The output from a BAG decoder is fed into six different TW decoders to better learn the features that are specific for a particular coefficient. The L1 distance between the network output and the ground truth is computed independently as the loss function for each coefficient. The BAG decoder contains 4 BAG convolutions with output channels of 256, 128, 64, 32, respectively. Each TW decoder block consists of one interpolation layer and two BAG convolutions with stride 1. 

\section{Experimental Results}
\label{sec:results}
\subsection{Dataset}
The Human Connectome Project (HCP) database\footnote{\href{https://www.humanconnectome.org/}{https://www.humanconnectome.org/}}~\cite{van2013wu} includes anatomical T1-weighted imaging and diffusion-weighted imaging acquired using a 3T Siemens ‘Connectom’ Skyra scanner. The high resolution T1-weighted data were acquired with 0.7 mm isotropic resolution, TR/TE = 2400/2.14 ms, and flip angle = 8°. The high-resolution diffusion MRI data were acquired with 1.25 mm isotropic resolution, TR/TE = 5520/89.5 ms, and flip angle = 78°. The diffusion MRI protocol consists of three diffusion-weighted shells which refers to b-value of 1000, 2000, and 3000. Each shell consists of 90 directions. 100 different subjects were selected from the HCP database in our study. Bias correction was applied to the structural T1 images~\cite{tustison2010n4itk}. The corrected image was then registered to the standard MNI space~\cite{evans19933d} using linear (FLIRT) and non-linear (FNIRT) registration tools from the FSL package~\cite{jenkinson2012fsl}. The preprocessing steps of the raw diffusion images included corrections for motion, susceptibility distortions, gradient non-linearity and eddy currents~\cite{andersson2015non,andersson2016integrated}. The processed diffusion data was then subsampled to 32 directions b1000 DWI to represent as a clinical accessible single-shell low-angular-resolution data~\cite{zeng2022fod}. FreeSurfer~\cite{fischl2012freesurfer} was used to generate the final brain mask for each subject~\cite{glasser2013minimal} and the corresponding diffusion tensor model was fit at each voxel~\cite{behrens2003characterization} using DTIFIT from FSL to generate the corresponding DTI volumes. All the processed images have a size of $6 \times 145 \times 174 \times 145$.

\subsection{Experimental Settings}
We split the 100 preprocessed DTIs into a training and testing set with a commonly used 80-20\% ratio, respectively. Each tensor coefficient from training set was z-score normalized before being cropped into patches. Two different patch sizes including $64 \times 64 \times 32$ with a $32 \times 32 \times 16$ overlap and $64 \times 64 \times 64$ with a $32 \times 32 \times 32$ overlap were set to conduct the efficiency study. We discarded any patches containing only the background region to reduce the computational cost in regions ignored by our loss function during training. All the models were trained for 50 epochs using Adam with an initial learning rate of 0.001 and the L1 loss inside the brain for batches of 4 patches on a single NVIDIA GeForce Tesla V100-SXM2 GPU. For inference, we reconstruct the final DTI volume by concatenating the predictions of all the cropped patches. The code was implemented on pytorch (version 1.10.1) and numpy package (version 1.21.2).

\subsection{Quantitative Results}
In general, the magnitudes of the tensor coefficients are small, with mean minimum and maximum values of -0.0044 and 0.0045, respectively. Hence the mean squared error (MSE) is omitted and only the mean absolute error (MAE) and peak signal noise ratio (PSNR) between inpainted and ground truth DTIs were reported as shown in Table~\ref{table_cv}. The proposed TW-BAG network achieved the best performance on this metric, which reduced the MAE of the tensor coefficients from 0.0041 to only 0.0009 and increased the PSNR by more than 10 dB. In addition to the traditional computer vision metrics, we also introduce common scalar metrics derived from DTIs to further evaluate the effect of inpainting with respect to micro-structural information in the disrupted regions. The tensor model can be characterized using the three main eigenvalues ($\lambda_1$, $\lambda_2$, and $\lambda_3$) of the 3 $\times$ 3 tensor matrix, ordered from the largest value to the lowest one. Several different metrics based on these eigenvalues have been defined but here we compute the most commonly used ones to characterize water diffusivity (AD, MD, and FA) that are defined as:

\begin{equation}
    AD = \lambda_1
\end{equation}

\begin{equation}
    MD = \frac{\lambda_1 + \lambda_2 + \lambda_3}{3}
\end{equation}

\begin{equation}
    FA = \sqrt{\frac{1}{2}} \frac{\sqrt{(\lambda_1-\lambda_2)^2 + (\lambda_1-\lambda_3)^2 + (\lambda_2-\lambda_3)^2 }}{\sqrt{\lambda_1^2 + \lambda_2^2 + \lambda_3^2}}
\end{equation}

The MAE of AD and MD between cropped image and ground truth in the disrupted region was 0.0013 and 0.0014, respectively. All the inpainting methods were able to reduce the MAE to less than 0.0001, hence the results of these two scalar metrics are omitted from our analysis. The MAE of cropped region and whole-brain FA between the predictions of the different compared methods and the ground truth are reported in Table~\ref{table_scalar}. In that setting, the proposed TW-BAG network also achieved the best performance by reducing the regional FA to 0.0327 (from 0.1888) and the whole-brain FA to 0.0018 (from 0.0105).

\begin{table}[htbp]
\caption{MAE and PSNR between the ground truth and inpainted results among different methods}
\begin{center}
\begin{tabular}{c|c|c}
\hline
Methods & MAE $\downarrow$ & PSNR $\uparrow$ \\ \ChangeRT{1pt}
Cropped & 0.0041$\pm$0.0004 & 51.8323$\pm$0.9006 \\ \hline
U-VQVAE~\cite{ayub2020inpainting} & 0.0015$\pm$0.0001 & 58.7779$\pm$0.6477 \\ \hline
w/o BAG-TW & 0.0013$\pm$0.0001 & 60.0626$\pm$0.6098 \\ 
w/o BA-TW & 0.0013$\pm$0.0001 & 60.0557$\pm$0.5986   \\ 
w/o TW & 0.0010$\pm$0.0001 & 62.3824$\pm$0.7312  \\ 
\textbf{TW-BAG} & \textbf{0.0009$\pm$0.0001} & \textbf{62.4051$\pm$0.6069}  \\ 
\hline
\end{tabular}
\label{table_cv}
\end{center}
\end{table}

\begin{table}[!htb]
\caption{MAE of FA and whole-brain FA between the ground truth and inpainted results among different methods}
\begin{center}
\begin{tabular}{c|c|c}
\hline
Methods & FA $\downarrow$ & Whole-Brain FA $\downarrow$ \\ \ChangeRT{1pt}
Cropped & 0.1888$\pm$0.0212 & 0.0105$\pm$0.0025\\ \hline
U-VQVAE & 0.0949$\pm$0.0372 & 0.0054$\pm$0.0026 \\ \hline
w/o BAG-TW & 0.0801$\pm$0.0399 & 0.0046$\pm$0.0027 \\ 
w/o BA-TW & 0.0820$\pm$0.0385 & 0.0047$\pm$0.0026 \\
w/o TW & 0.0732$\pm$0.0204 & 0.0041$\pm$0.0015 \\ 
\textbf{TW-BAG} & \textbf{0.0327$\pm$0.0192} & \textbf{0.0018$\pm$0.0012}  \\ 
\hline
\end{tabular}
\label{table_scalar}
\end{center}
\end{table}

\subsection{Qualitative Results}
\begin{figure*}[!hth]
\centering
\begin{tabular}{ccccccc}
\includegraphics[width=0.12\textwidth]{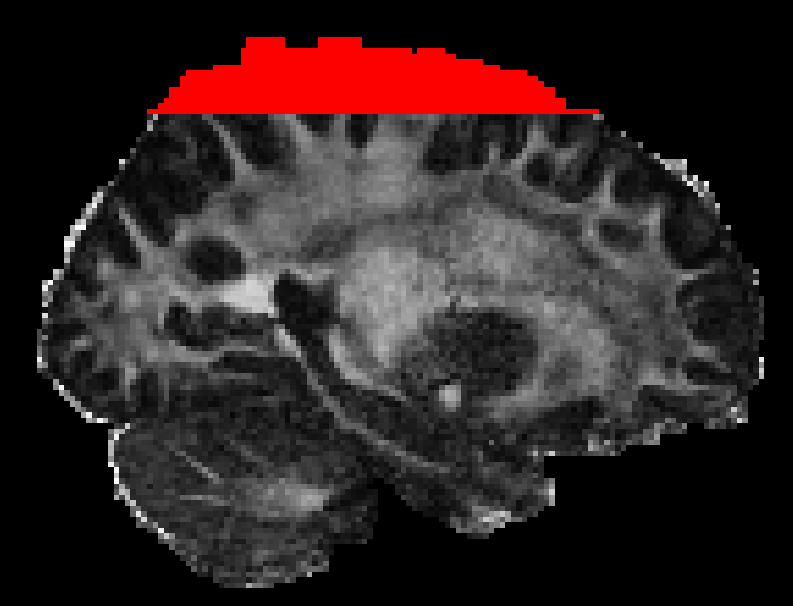} &
\includegraphics[width=0.12\textwidth]{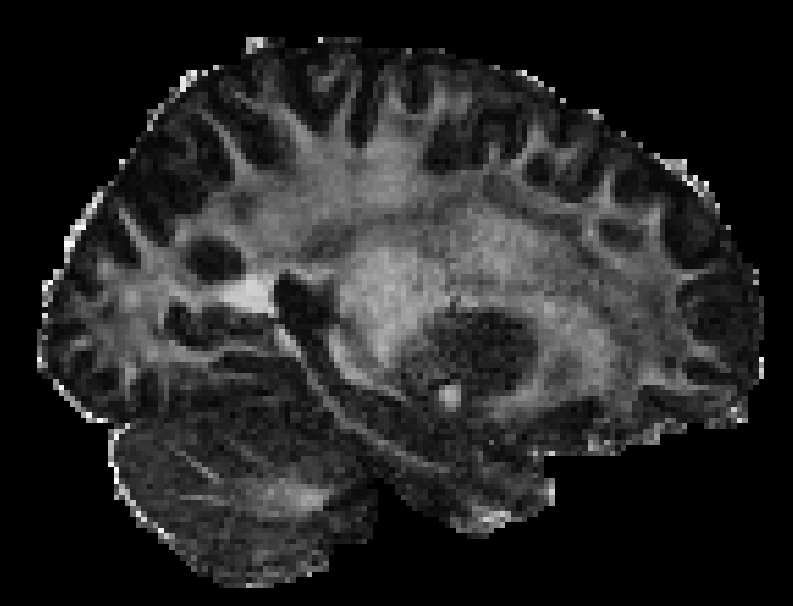} &
\includegraphics[width=0.12\textwidth]{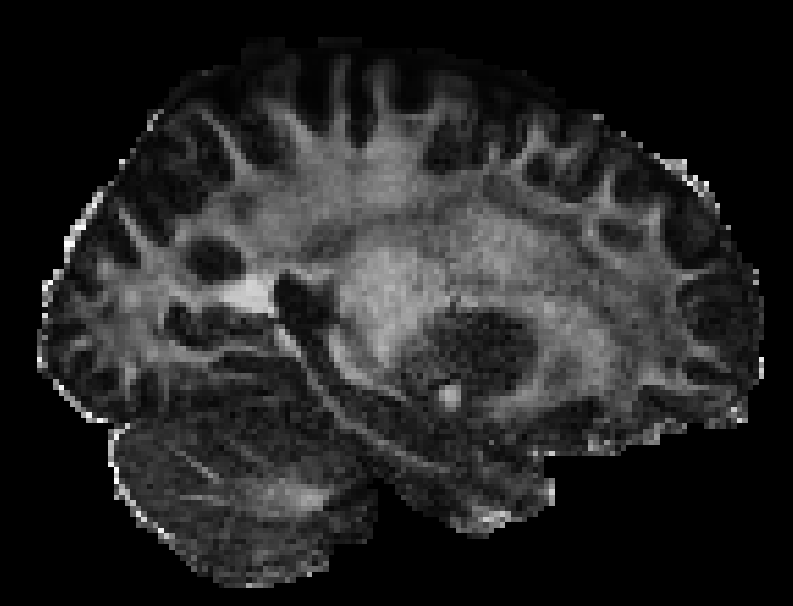} &
\includegraphics[width=0.12\textwidth]{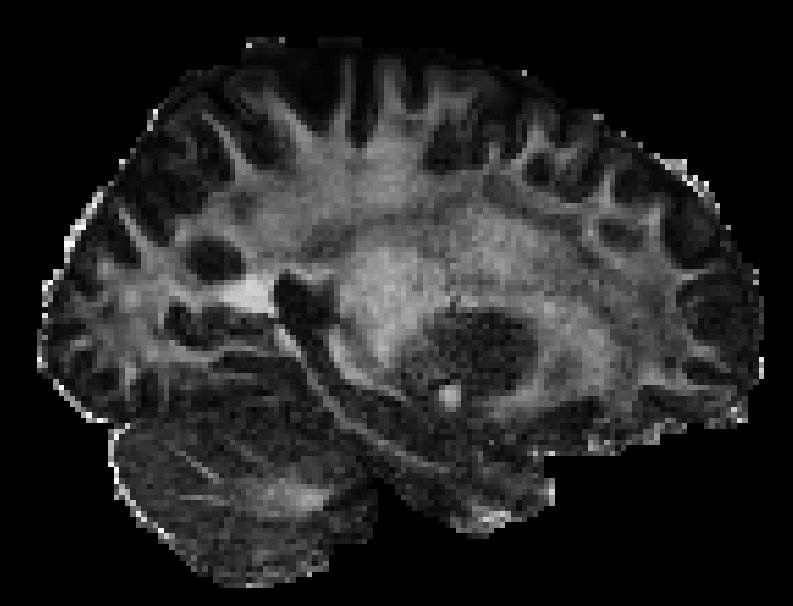} &
\includegraphics[width=0.12\textwidth]{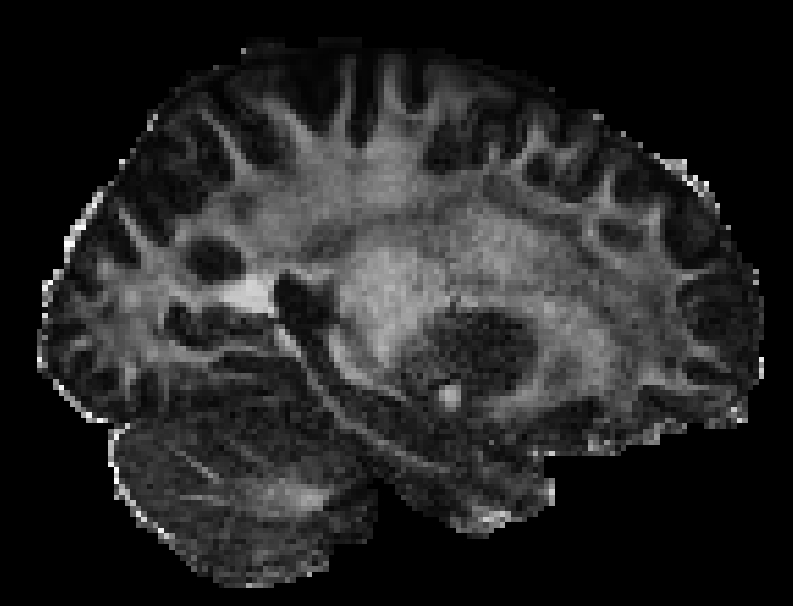} &
\includegraphics[width=0.12\textwidth]{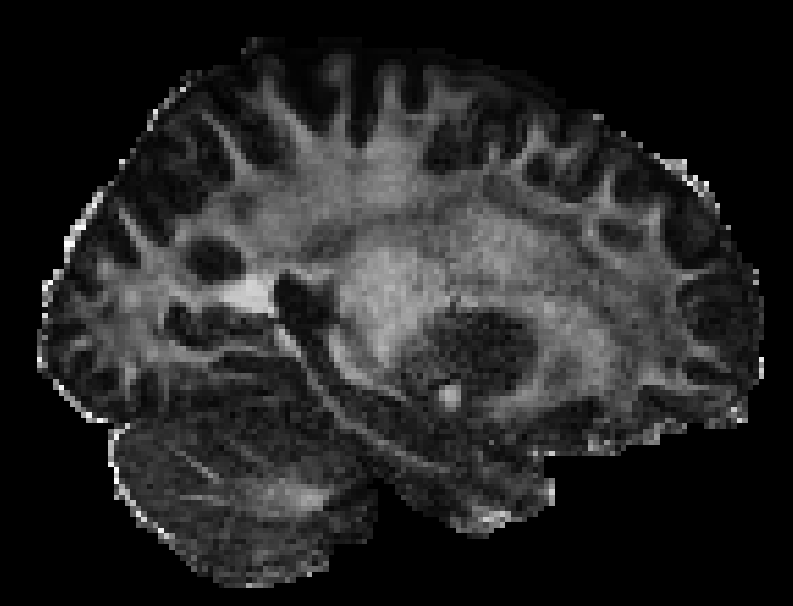} &
\includegraphics[width=0.12\textwidth]{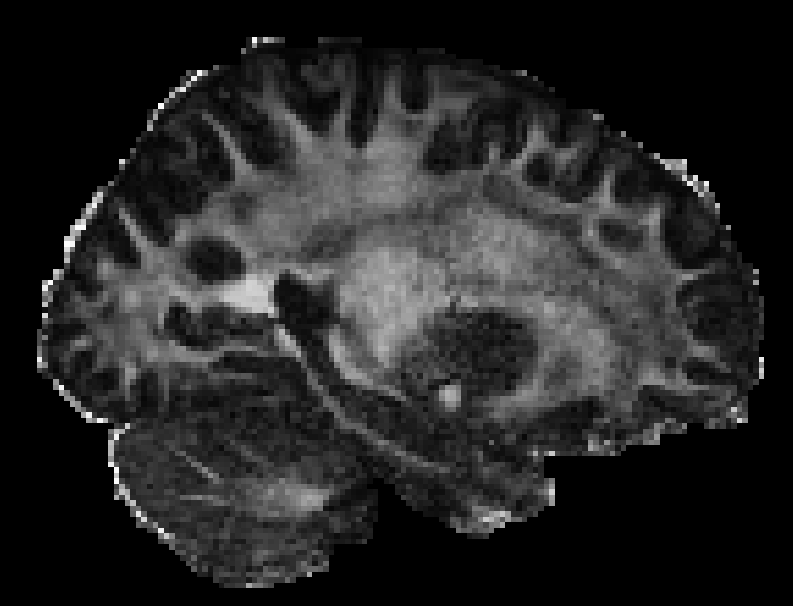}\\
\end{tabular}
\begin{tabular}{ccccccc}
\includegraphics[width=0.12\textwidth]{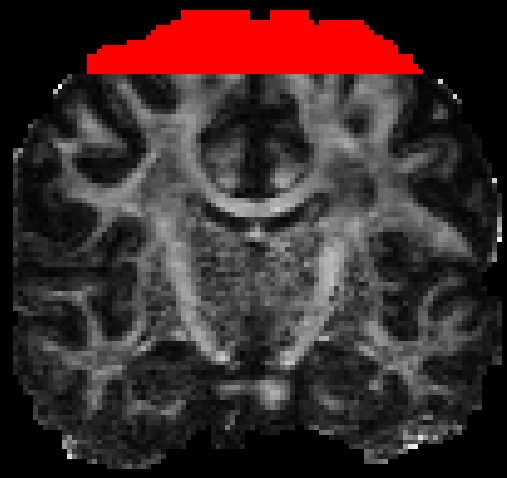} &
\includegraphics[width=0.12\textwidth]{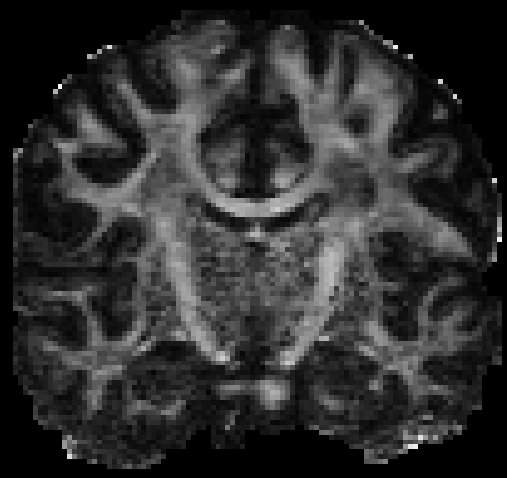} &
\includegraphics[width=0.12\textwidth]{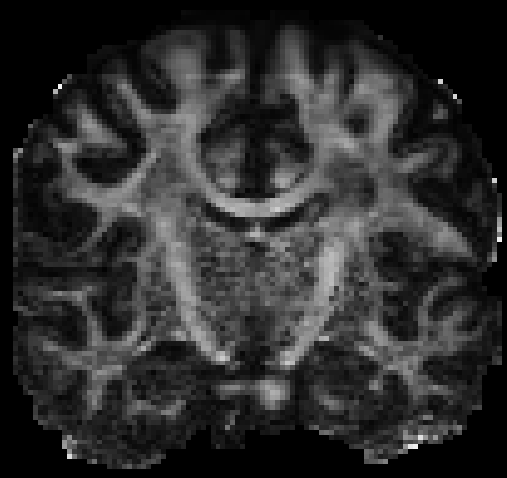} &
\includegraphics[width=0.12\textwidth]{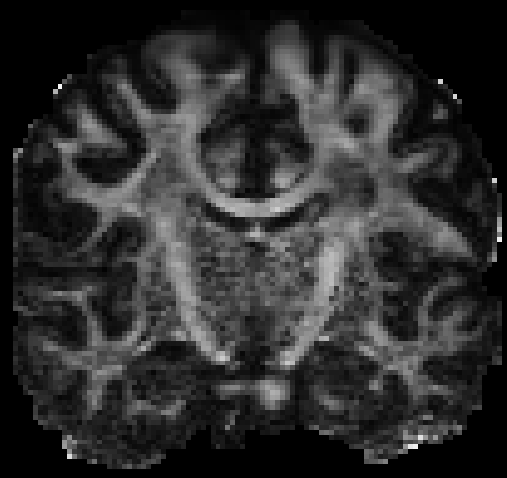} &
\includegraphics[width=0.12\textwidth]{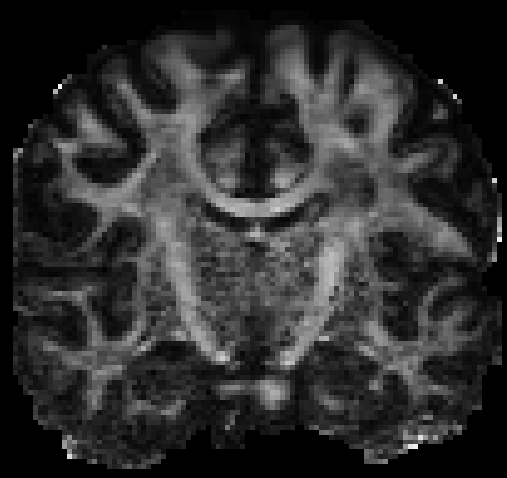} &
\includegraphics[width=0.12\textwidth]{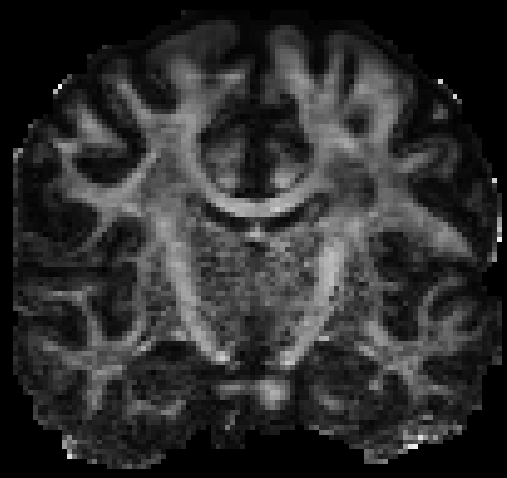} &
\includegraphics[width=0.12\textwidth]{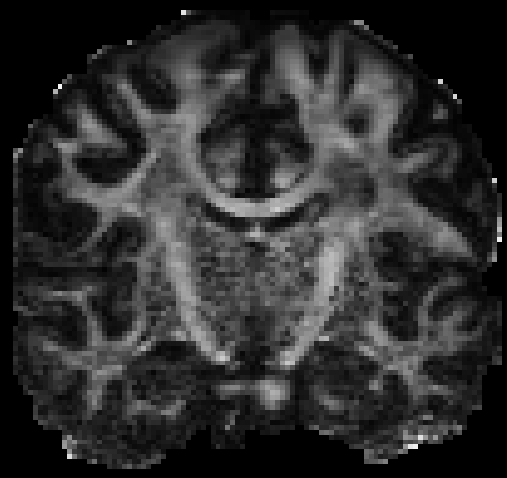}\\
\end{tabular}
\begin{tabular}{ccccccc}
\includegraphics[width=0.12\textwidth]{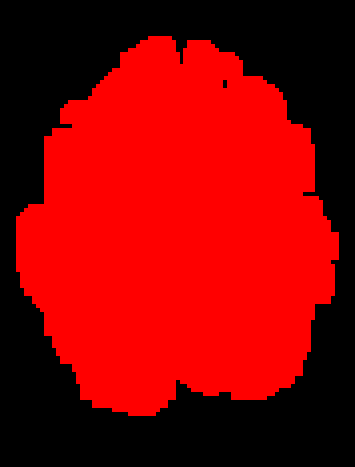} &
\includegraphics[width=0.12\textwidth]{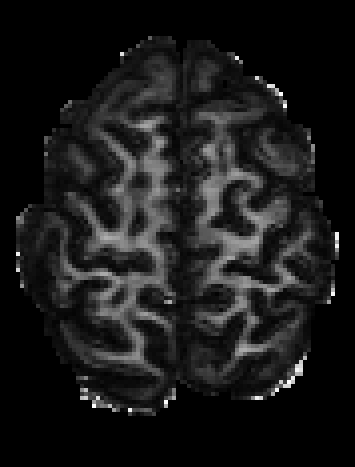} &
\includegraphics[width=0.12\textwidth]{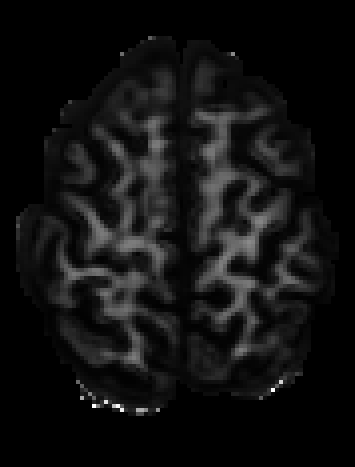} &
\includegraphics[width=0.12\textwidth]{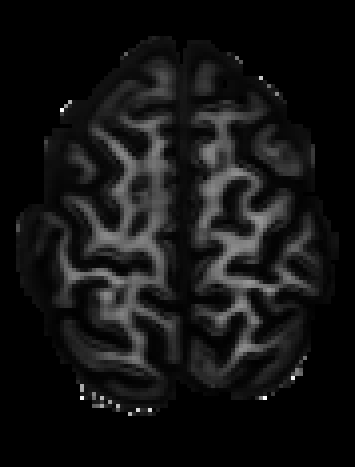} &
\includegraphics[width=0.12\textwidth]{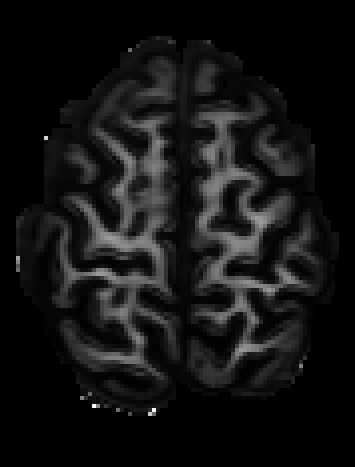} &
\includegraphics[width=0.12\textwidth]{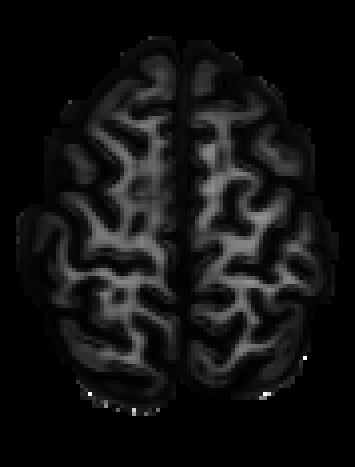} &
\includegraphics[width=0.12\textwidth]{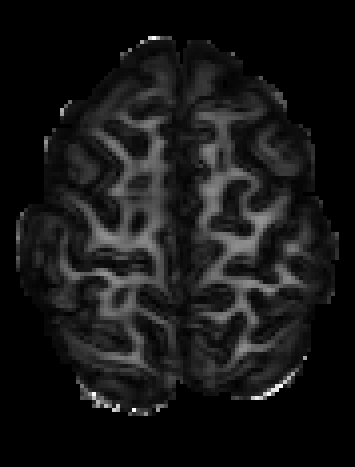}\\
\small (a) Disrupted & \small (b) GT & \small (c) U-VQVAE & \small (d) Baseline & \small (e) w/o BA-TW & \small (f) w/o TW & \textbf{(g) TW-BAG} \\[6pt]
\end{tabular}
\caption{FA maps derived from the results obtained by the compared methods with (a) as the input of the network with the values in the disrupted region (marked in red) set to 0. (b) is the ground truth image, (c)-(g) are the corresponding FA maps derived from inpainted DTIs generated by the U-VQVAE, our proposed method without BAG-TW, BA-TW, TW and the proposed TW-BAG, respectively.}
\label{fig:vis_fa}
\end{figure*}

The visualization of inpainted results (showing the whole diffusion tensor for each voxel) generated from different models are demonstrated in Figure~\ref{fig:vis_tensor}, where the tensor is represented in RGB coding that colours red, green and blue represent diffusion in the x, y, and z axes respectively. One typical example of reconstruction in the disrupted ROIs has been zoomed for visualization. Compared to the ground truth, TW-BAG is able to generate more distinct orientations following the original distribution and intensity across sagittal, coronal and axial views, which could demonstrate water diffusivity (denoted by eigenvalues of tensor model) more accurately. As indicated in Figure~\ref{fig:vis_fa}, the boundaries and textures of the FA map for TW-BAG are less blurry than the other competing models, validating the effectiveness of TW-BAG from a neuroscience perspective.  

\begin{figure}[htbp]
\includegraphics[width=\columnwidth]{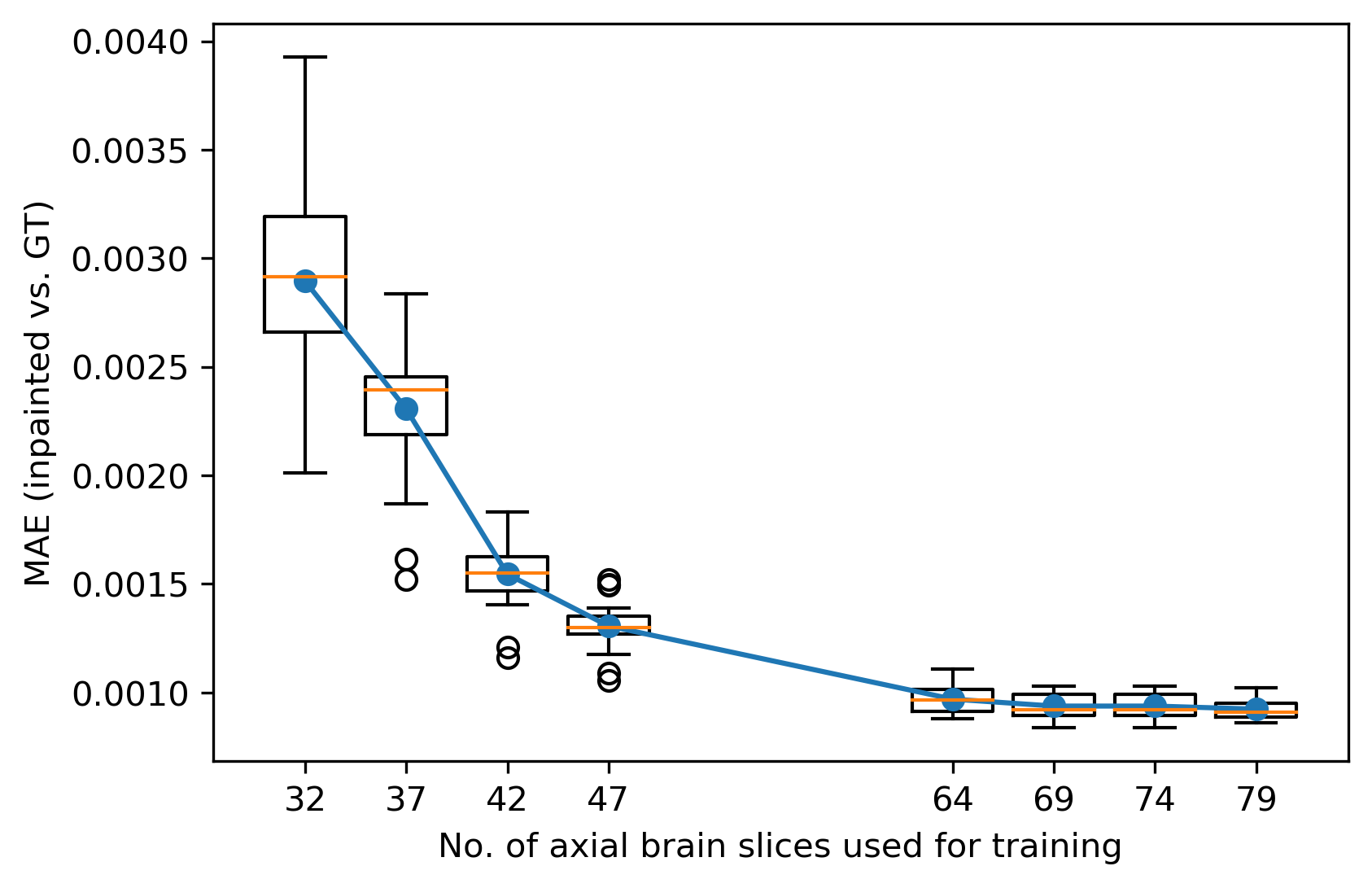}
\caption{MAE between the ground truth and inpainted tensors using different percentages of training brain slices.}
\label{fig:box_mae}
\end{figure}

\begin{figure}[htbp]
\includegraphics[width=\columnwidth]{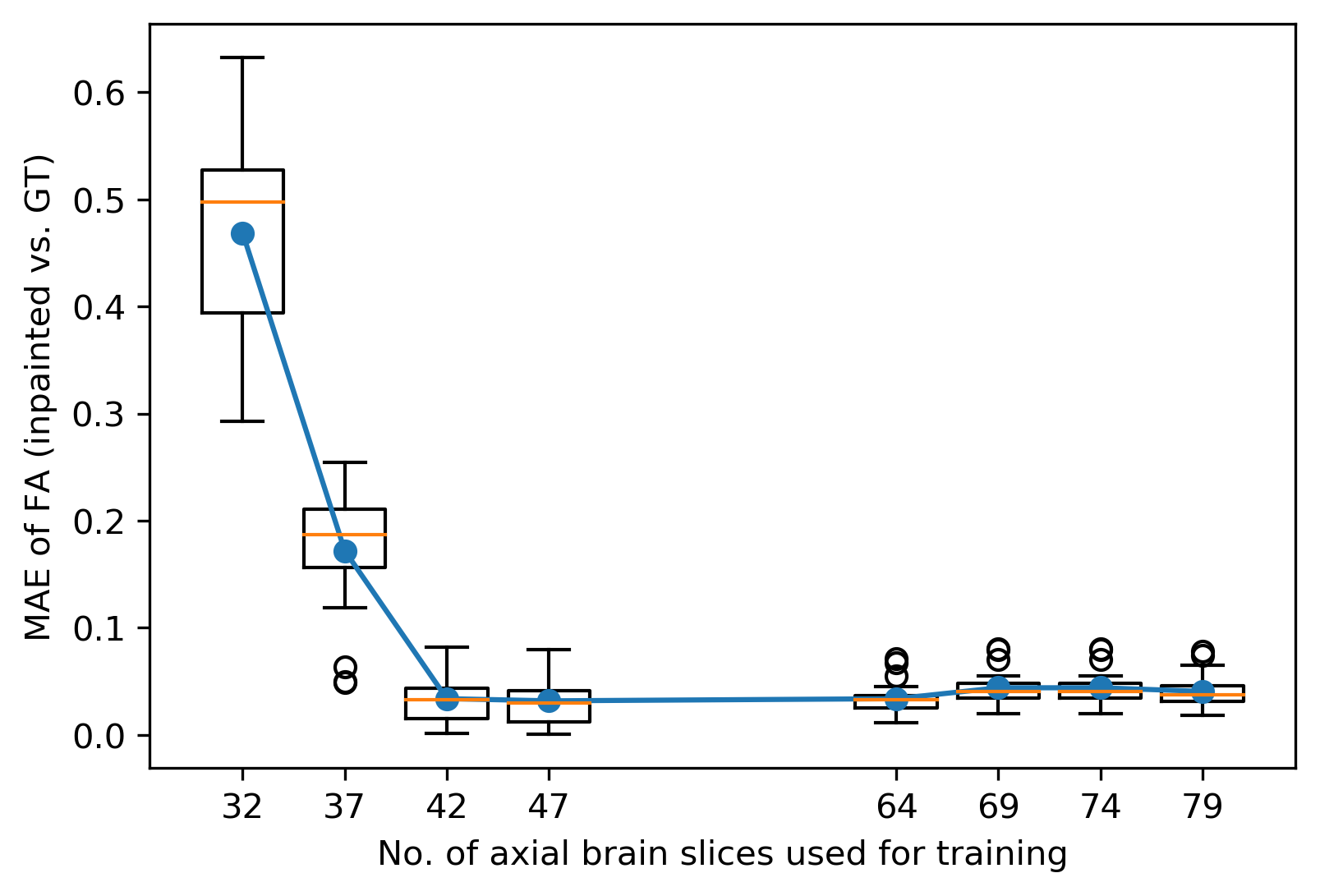}
\caption{MAE between the FA derived from the ground truth and inpainted tensors using different percentages of training brain slices.}
\label{fig:box_fa}
\end{figure}

\subsection{Ablation Study}
To better understand each new component from our proposal, an ablation study was conducted for the gate convolution, brain-aware mechanism and tensor-wise coefficient specific decoder separately. Summaries of the metrics presented in the previous section for each ablation are shown in Table~\ref{table_cv} and Table~\ref{table_scalar}, respectively. The baseline model (w/o BAG-TW) was designed as a 3D-UNet and all the experiments included were built on top of that baseline structure. The gate convolution (w/o BA-TW) reached a slightly worse result when compared to the baseline, while the brain-aware mechanism (w/o TW) increased the performance, especially on the scalar metrics. The results suggest that the gate convolution failed to focus on extracting the features from only the valid brain and the brain-aware mechanism tackled this dilemma by introducing a guidance on the foreground. Although the introduction of specifically designed tensor-wise coefficient decoders only increased the MAE and PSNR by a small margin, it decreased the MAE for FA from 0.07 to 0.03 (55\% improvement) and the whole-brain FA from 0.0041 to 0.0018 (56\% improvement). The scalar metrics prove the effectiveness of the tensor-wise decoders in learning features tailored for each particular coefficient in the tensor model.

\subsection{Efficiency Study}
\begin{figure*}[!htb]
\centering

\begin{tabular}{cccccc}
\includegraphics[width=0.14\textwidth]{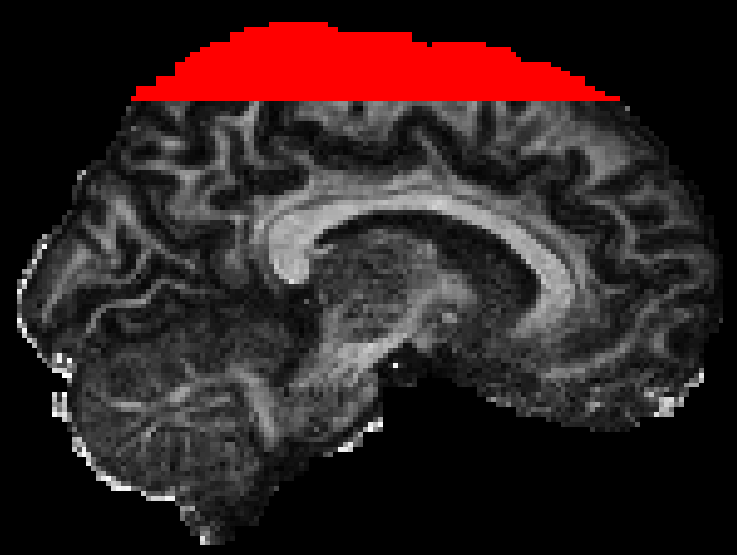} &
\includegraphics[width=0.14\textwidth]{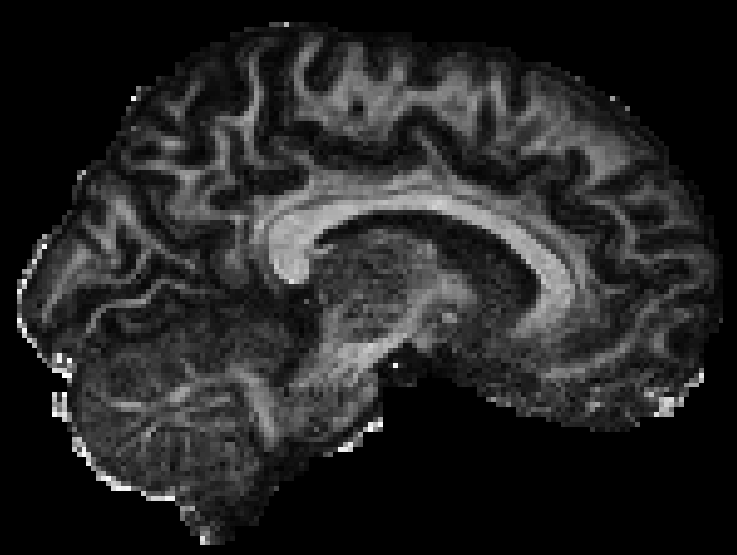} &
\includegraphics[width=0.14\textwidth]{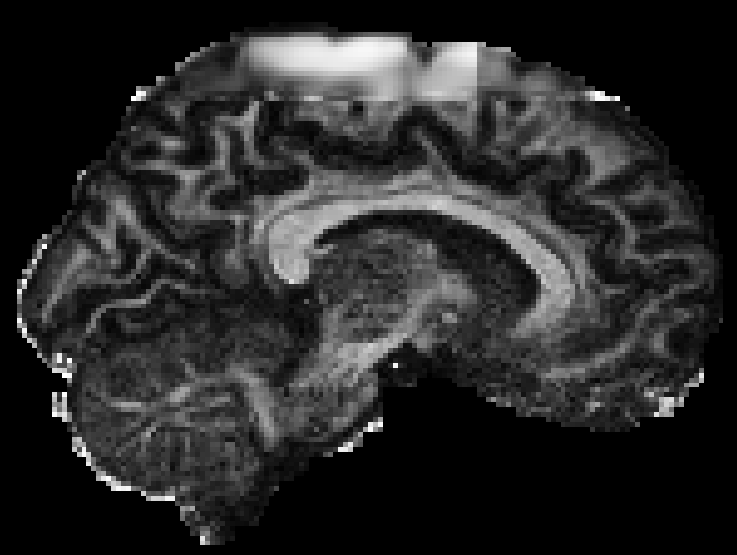} &
\includegraphics[width=0.14\textwidth]{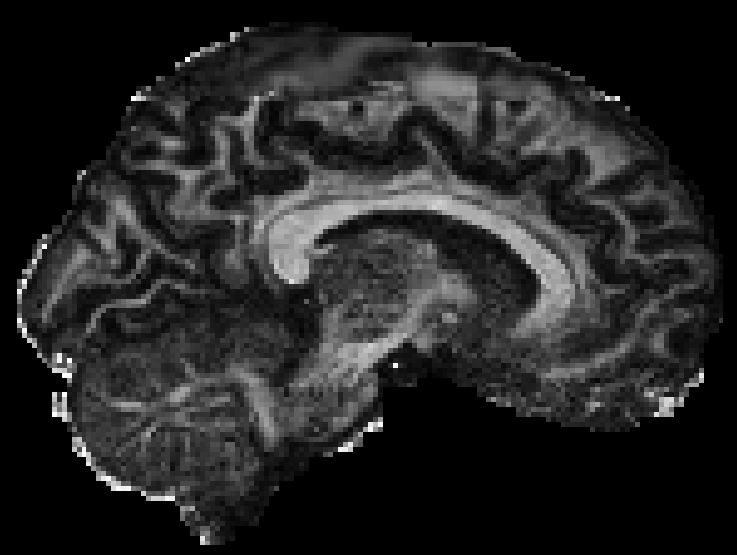} & 
\includegraphics[width=0.14\textwidth]{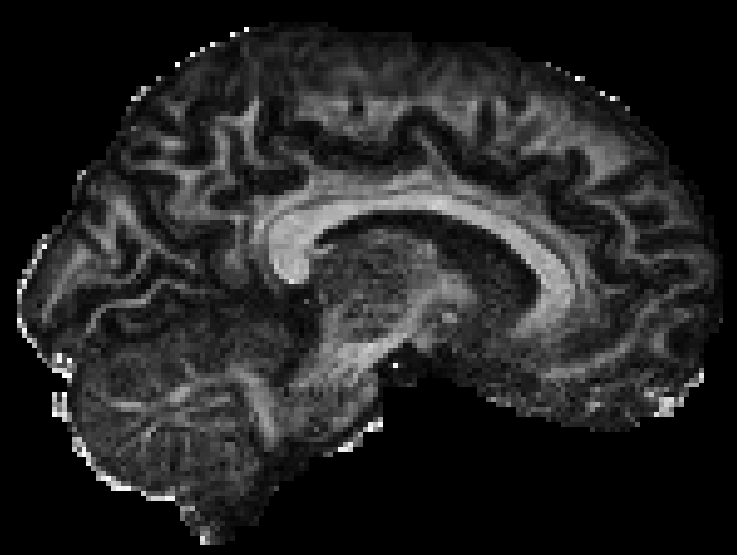} &
\includegraphics[width=0.14\textwidth]{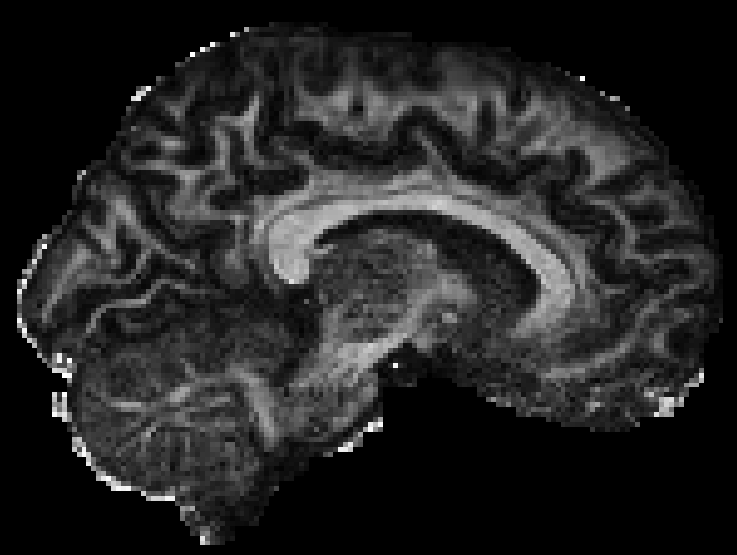}\\
\end{tabular}

\begin{tabular}{cccccc}
\includegraphics[width=0.14\textwidth]{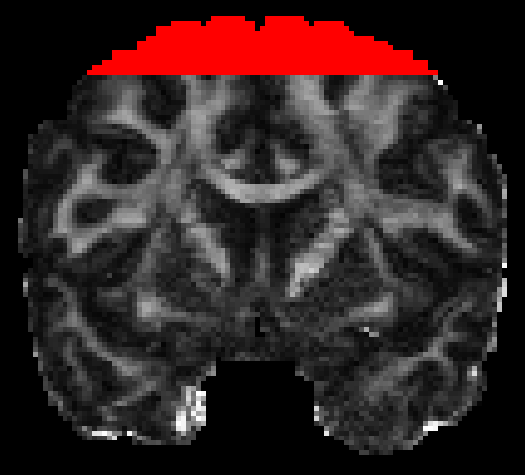} &
\includegraphics[width=0.14\textwidth]{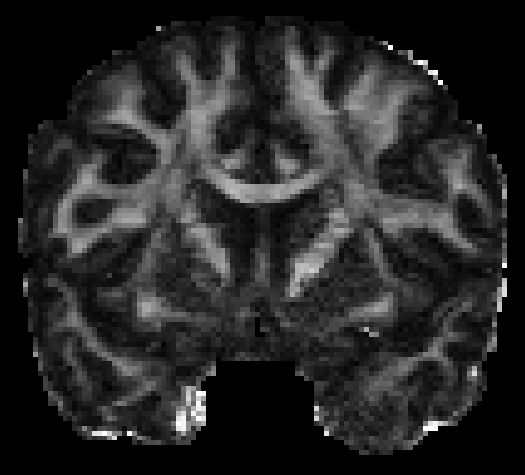} &
\includegraphics[width=0.14\textwidth]{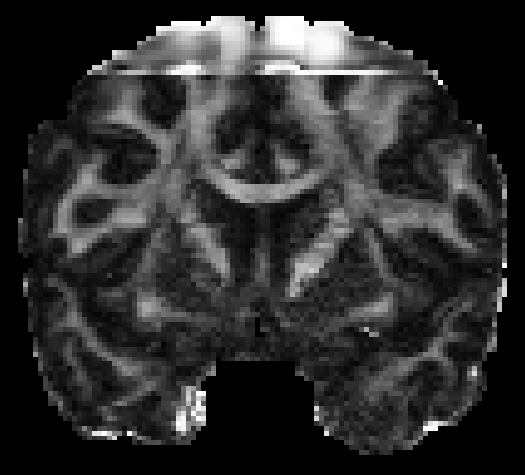} &
\includegraphics[width=0.14\textwidth]{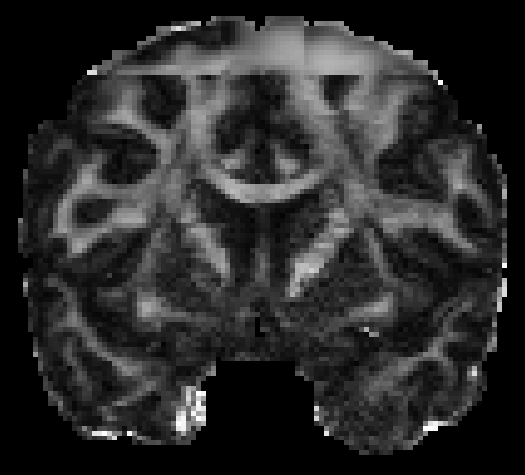} & 
\includegraphics[width=0.14\textwidth]{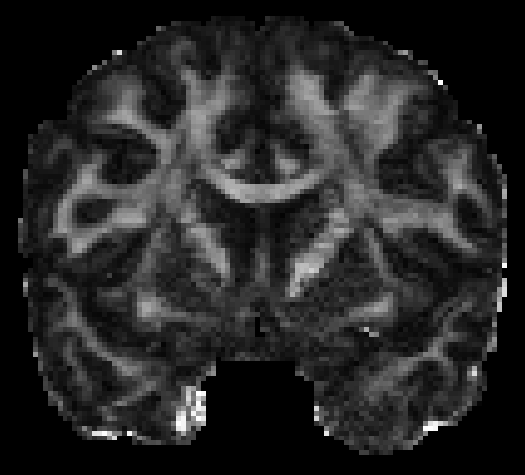} &
\includegraphics[width=0.14\textwidth]{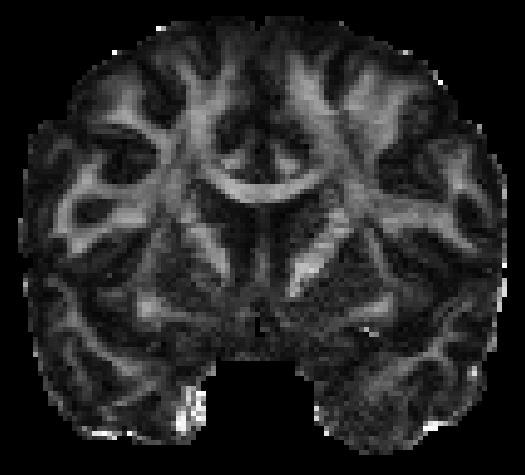}\\
\end{tabular}

\begin{tabular}{cccccc}
\includegraphics[width=0.14\textwidth]{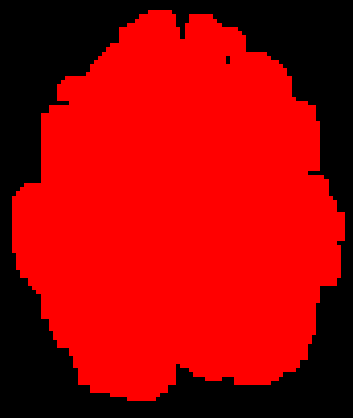} &
\includegraphics[width=0.14\textwidth]{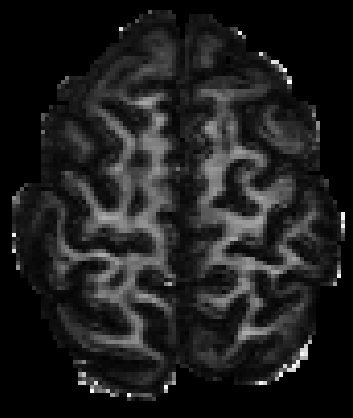} &
\includegraphics[width=0.14\textwidth]{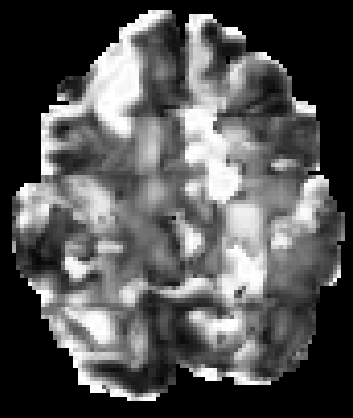} &
\includegraphics[width=0.14\textwidth]{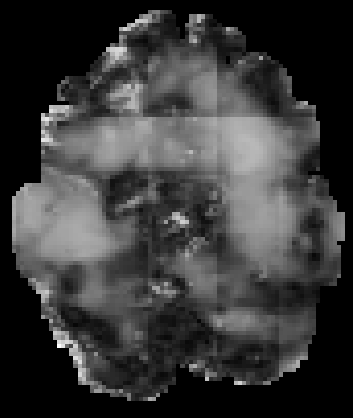} & 
\includegraphics[width=0.14\textwidth]{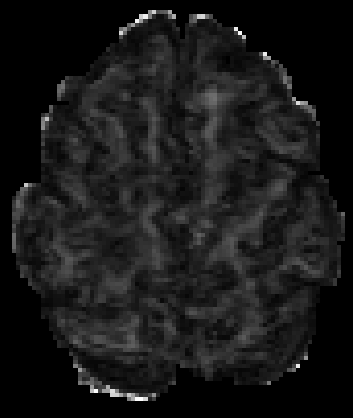} &
\includegraphics[width=0.14\textwidth]{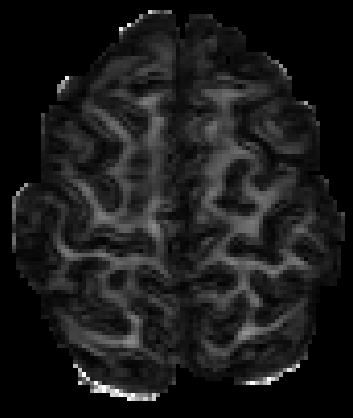}\\
\small (a) Disrupted & \small (b) GT & \small (c) 32 & \small (d) 37 & \small (e) 42 & \small (f) 47 \\[6pt]
\end{tabular}
\caption{FA maps derived from the results obtained by the compared methods with (a) as the input of the network with the values in the disrupted region (marked in red) set to 0. (b) is the ground truth image, (c)-(f) are the corresponding FA maps derived from inpainted DTIs generated by training with 32, 37, 42, 47 valid brain slices, respectively.}
\label{fig:vis_efficay}
\end{figure*}

Due to the fact that brain tissues are continuous and follow a common structure, we decided to explore how much valid brain information is sufficient to reconstruct the disrupted regions using the same metrics as the previous sections. The same two sets of patch sizes were included in the efficiency study. Due to the limitation of needing at least one missing slice per patch to learn meaningful information, the maximum number of possible additional brain slices was 47 for the smaller patches ($64 \times 64 \times 32$). Quantitative results are shown in Figure~\ref{fig:box_mae} and Figure~\ref{fig:box_fa} while qualitative examples for the FA maps in three anatomical planes are illustrated in Figure~\ref{fig:vis_efficay}. When constraining the number of brain slices to 32, the results reach the largest MAE. As shown in Figure~\ref{fig:vis_efficay} (b), the input brain slices are not sufficient for the network to learn the information of brain morphology due to the limited amount of patient's information (roughly half the patch contains missing slices). As the number of possible additional training patches was increased, more morphological information was learned by the network and further detail was observed in terms of anatomy as shown in Figure~\ref{fig:vis_efficay} (c) (d) (e). With larger training patches ($64 \times 64 \times 64$), the MAE between ground truth and inpainted DTIs was reduced significantly. Although a larger range of valid patches with at least 1 missing slice is possible with a larger patch size, the performance increase was marginal (visual samples are not shown due to minimal differences). According to Figure~\ref{fig:box_fa}, the MAE for FA decreases when the number of valid brain patches increases and reaches a minimum with the 47 limit for the smaller patches. However, there seems to be no significant improvement in terms of FA by further increasing the number of brain slices for training. The experimental results show that 47 ($\sim$28\% of the whole brain) additional brain slices is sufficient to reconstruct the missing information in the 15 ($\sim$10\% of the whole brain) disrupted slices.




\subsection{Clinical Impact}
Clinical studies have shown that decreased whole-brain FA associates with ageing related neurodegeneration~\cite{kochunov2007relationship}, while a significant reduction of FA in the brain is commonly linked to a significant cognition impairment~\cite{sexton2011meta}. Studies have also shown FA differences between age and gender matched healthy controls and patients with various brain diseases. To illustrate the clinical impact of the FA error due to cropping, FA differences between controls and different disease groups including Multiple Sclerosis (MS), Alzheimer's Disease (AD), Amyotrophic Lateral Sclerosis (ALS), Parkinson's Disease–Mild Cognitive Impairment (PD-MCI) and Parkinson's Disease-Dementia (PD-D), are provided in Table~\ref{table_clinical}. For example, a 0.03 whole brain FA difference was observed between healthy controls and MS patients~\cite{cercignani2001mean}. Similarly, decreased FA values were also observed in white matter (AD, ALS, PD-MCI, PD-D) and grey matter (AD and ALS) for different patient groups~\cite{giulietti2018whole, rose2008gray, metwalli2010utility, melzer2013white}. The synthesized cropped brains used in our experiments had a whole-brain FA MAE of 0.01, which could potentially impact the observed group differences between healthy controls and patients. Regarding the mean FA in pathological ROIs, a difference of 0.16 was found between MS lesions and normal appearing white matter regions~\cite{schmierer2007diffusion}, which is once again close to the MAE caused by the missing brain slices. Our proposed approach reduces the FA errors in the cropped region and the whole brain by 0.1561 and 0.0087, respectively. Consequently, these measurements errors are lower than the gap between controls and patients, avoiding erroneous conclusions in clinical studies and trials. Furthermore, the subjects with disrupted DTIs that are commonly discarded could now be included in these studies by inpainting the missing information and mitigate the impact of these regions in terms of FA metrics.

\begin{table}[!htb]
\caption{FA reduction of patient groups from different disease types compared to controls}
\begin{center}
\begin{tabular}{c|c|c}
\hline
Disease type & Mean FA reduction & ROI\\ \ChangeRT{1pt}
\multicolumn{3}{c}{Pathology} \\ \ChangeRT{1pt}
MS & 0.16 & MS lesions \\ \ChangeRT{1pt}
\multicolumn{3}{c}{Whole-brain} \\ \ChangeRT{1pt}
MS & 0.03 & Whole brain \\ \hline
AD & 0.01 & Whiter matter \\
AD & 0.06 & Grey matter \\ \hline
ALS & 0.02 & Whiter matter \\
ALS & 0.05 & Grey matter \\ \hline
PD-MCI & 0.02 & Whiter matter \\
PD-D & 0.03 & Whiter matter \\ \hline
\end{tabular}
\label{table_clinical}
\end{center}
\end{table}

\section{Conclusion}
\label{sec:conclusions}
We have proposed a network called TW-BAG specifically tailored to inpainting cropped DTIs due to sub-optimal acquisition settings. The proposed approach achieved superior performance on the restoration of the missing slices with respect to traditional computer vision metrics and scalar diffusion metrics commonly used in clinical studies. Our experimental results show that TW-BAG mitigates the impact introduced by disrupted regions by inpainting the missing clinical information using the information of the rest of the brain. Therefore, TW-BAG can be used as a post-scanning process to address the disrupted diffusion regions without discarding valuable scans for clinical studies and it can be easily extended to inpaint other disrupted ROIs.




\bibliographystyle{IEEEtran}
\bibliography{bibliography.bib}

\begin{thebibliography}{10}
\providecommand{\url}[1]{#1}
\csname url@samestyle\endcsname
\providecommand{\newblock}{\relax}
\providecommand{\bibinfo}[2]{#2}
\providecommand{\BIBentrySTDinterwordspacing}{\spaceskip=0pt\relax}
\providecommand{\BIBentryALTinterwordstretchfactor}{4}
\providecommand{\BIBentryALTinterwordspacing}{\spaceskip=\fontdimen2\font plus
\BIBentryALTinterwordstretchfactor\fontdimen3\font minus
  \fontdimen4\font\relax}
\providecommand{\BIBforeignlanguage}[2]{{%
\expandafter\ifx\csname l@#1\endcsname\relax
\typeout{** WARNING: IEEEtran.bst: No hyphenation pattern has been}%
\typeout{** loaded for the language `#1'. Using the pattern for}%
\typeout{** the default language instead.}%
\else
\language=\csname l@#1\endcsname
\fi
#2}}
\providecommand{\BIBdecl}{\relax}
\BIBdecl

\bibitem{baliyan2016diffusion}
V.~Baliyan, C.~J. Das, R.~Sharma, and A.~K. Gupta, ``Diffusion weighted
  imaging: technique and applications,'' \emph{World journal of radiology},
  vol.~8, no.~9, p. 785, 2016.

\bibitem{soares2013hitchhiker}
J.~M. Soares, P.~Marques, V.~Alves, and N.~Sousa, ``A hitchhiker's guide to
  diffusion tensor imaging,'' \emph{Frontiers in neuroscience}, vol.~7, p.~31,
  2013.

\bibitem{feldman2010diffusion}
H.~M. Feldman, J.~D. Yeatman, E.~S. Lee, L.~H. Barde, and S.~Gaman-Bean,
  ``Diffusion tensor imaging: a review for pediatric researchers and
  clinicians,'' \emph{Journal of developmental and behavioral pediatrics:
  JDBP}, vol.~31, no.~4, p. 346, 2010.

\bibitem{ma2022multiple}
Y.~Ma, C.~Zhang, M.~Cabezas, Y.~Song, Z.~Tang, D.~Liu, W.~Cai, M.~Barnett, and
  C.~Wang, ``Multiple sclerosis lesion analysis in brain magnetic resonance
  images: techniques and clinical applications,'' \emph{IEEE Journal of
  Biomedical and Health Informatics}, vol.~26, no.~6, pp. 2680--2692, 2022.

\bibitem{tae2018current}
W.~S. Tae, B.~J. Ham, S.~B. Pyun, S.~H. Kang, and B.~J. Kim, ``Current clinical
  applications of diffusion-tensor imaging in neurological disorders,''
  \emph{Journal of Clinical Neurology}, vol.~14, no.~2, pp. 129--140, 2018.

\bibitem{assaf2008diffusion}
Y.~Assaf and O.~Pasternak, ``Diffusion tensor imaging ({DTI})-based white
  matter mapping in brain research: a review,'' \emph{Journal of molecular
  neuroscience}, vol.~34, no.~1, pp. 51--61, 2008.

\bibitem{sbardella2013dti}
E.~Sbardella, F.~Tona, N.~Petsas, and P.~Pantano, ``{DTI} measurements in
  multiple sclerosis: evaluation of brain damage and clinical implications,''
  \emph{Multiple sclerosis international}, vol. 2013, 2013.

\bibitem{filippi2001diffusion}
M.~Filippi, M.~Cercignani, M.~Inglese, M.~Horsfield, and G.~Comi, ``Diffusion
  tensor magnetic resonance imaging in multiple sclerosis,'' \emph{Neurology},
  vol.~56, no.~3, pp. 304--311, 2001.

\bibitem{ciccarelli2001investigation}
O.~Ciccarelli, D.~Werring, C.~Wheeler-Kingshott, G.~Barker, G.~Parker,
  A.~Thompson, and D.~Miller, ``Investigation of {MS} normal-appearing brain
  using diffusion tensor {MRI} with clinical correlations,'' \emph{Neurology},
  vol.~56, no.~7, pp. 926--933, 2001.

\bibitem{duffau2014dangers}
H.~Duffau, ``The dangers of magnetic resonance imaging diffusion tensor
  tractography in brain surgery,'' \emph{World neurosurgery}, vol.~81, no.~1,
  pp. 56--58, 2014.

\bibitem{charlton2006white}
R.~A. Charlton, T.~Barrick, D.~McIntyre, Y.~Shen, M.~O'sullivan, F.~Howe,
  C.~Clark, R.~Morris, and H.~Markus, ``White matter damage on diffusion tensor
  imaging correlates with age-related cognitive decline,'' \emph{Neurology},
  vol.~66, no.~2, pp. 217--222, 2006.

\bibitem{wan2020bringing}
Z.~Wan, B.~Zhang, D.~Chen, P.~Zhang, D.~Chen, J.~Liao, and F.~Wen, ``Bringing
  old photos back to life,'' in \emph{Proceedings of the IEEE/CVF conference on
  computer vision and pattern recognition (CVPR)}, 2020, pp. 2747--2757.

\bibitem{yu2019free}
J.~{Yu}, Z.~{Lin}, J.~{Yang}, X.~{Shen}, X.~{Lu}, and T.~S. {Huang},
  ``Free-form image inpainting with gated convolution,'' in \emph{Proceedings
  of the IEEE International Conference on Computer Vision (ICCV)}, 2019, pp.
  4471--4480.

\bibitem{barnes2011patchmatch}
C.~{Barnes}, D.~B. {Goldman}, E.~{Shechtman}, and A.~{Finkelstein}, ``The
  patchmatch randomized matching algorithm for image manipulation,''
  \emph{Communications of the ACM}, vol.~54, no.~11, pp. 103--110, 2011.

\bibitem{ron2015unet}
O.~Ronneberger, P.~Fischer, and T.~Brox, ``U-{N}et: Convolutional networks for
  biomedical image segmentation,'' in \emph{International Conference on Medical
  Image Computing and Computer-Assisted Intervention (MICCAI)}.\hskip 1em plus
  0.5em minus 0.4em\relax Springer International Publishing, 2015, pp.
  234--241.

\bibitem{goodfellow2014generative}
I.~Goodfellow, J.~Pouget-Abadie, M.~Mirza, B.~Xu, D.~Warde-Farley, S.~Ozair,
  A.~Courville, and Y.~Bengio, ``Generative adversarial nets,'' \emph{Advances
  in neural information processing systems}, vol.~27, 2014.

\bibitem{creswell2018generative}
A.~Creswell, T.~White, V.~Dumoulin, K.~Arulkumaran, B.~Sengupta, and A.~A.
  Bharath, ``Generative adversarial networks: {A}n overview,'' \emph{IEEE
  signal processing magazine}, vol.~35, no.~1, pp. 53--65, 2018.

\bibitem{sal2021a}
N.~M.~F. Salem, ``A survey on various image inpainting techniques,''
  \emph{Future Engineering Journal}, vol.~2, no.~2, pp. 1/1--18, 2021.

\bibitem{mirza2014conditional}
M.~Mirza and S.~Osindero, ``Conditional generative adversarial nets,''
  \emph{arXiv preprint arXiv:1411.1784}, 2014.

\bibitem{isola2017image}
P.~Isola, J.-Y. Zhu, T.~Zhou, and A.~A. Efros, ``Image-to-image translation
  with conditional adversarial networks,'' in \emph{Proceedings of the IEEE/CVF
  conference on computer vision and pattern recognition (CVPR)}, 2017, pp.
  1125--1134.

\bibitem{zhang2018ms}
C.~Zhang, Y.~Song, S.~Liu, S.~Lill, C.~Wang, Z.~Tang, Y.~You, Y.~Gao,
  A.~Klistorner, M.~Barnett \emph{et~al.}, ``{MS-GAN}: {GAN}-based semantic
  segmentation of multiple sclerosis lesions in brain magnetic resonance
  imaging,'' in \emph{Digital Image Computing: Techniques and Applications
  (DICTA)}.\hskip 1em plus 0.5em minus 0.4em\relax IEEE, 2018, pp. 1--8.

\bibitem{liu2019nuclei}
D.~Liu, D.~Zhang, Y.~Song, C.~Zhang, F.~Zhang, L.~O'Donnell, and W.~Cai,
  ``Nuclei segmentation via a deep panoptic model with semantic feature
  fusion.'' in \emph{International Joint Conference on Artificial Intelligence
  (IJCAI)}, 2019, pp. 861--868.

\bibitem{jin2018ct}
D.~Jin, Z.~Xu, Y.~Tang, A.~P. Harrison, and D.~J. Mollura, ``C{T}-realistic
  lung nodule simulation from 3d conditional generative adversarial networks
  for robust lung segmentation,'' in \emph{International Conference on Medical
  Image Computing and Computer-Assisted Intervention (MICCAI)}.\hskip 1em plus
  0.5em minus 0.4em\relax Springer, 2018, pp. 732--740.

\bibitem{battaglini2012evaluating}
M.~{Battaglini}, M.~{Jenkinson}, and N.~D. {Stefano}, ``Evaluating and reducing
  the impact of white matter lesions on brain volume measurements,''
  \emph{Human brain mapping}, vol.~33, no.~9, pp. 2062--2071, 2012.

\bibitem{rohde2004comprehensive}
G.~K. Rohde, A.~Barnett, P.~Basser, S.~Marenco, and C.~Pierpaoli,
  ``Comprehensive approach for correction of motion and distortion in
  diffusion-weighted mri,'' \emph{Magnetic Resonance in Medicine: An Official
  Journal of the International Society for Magnetic Resonance in Medicine},
  vol.~51, no.~1, pp. 103--114, 2004.

\bibitem{ayub2020inpainting}
R.~Ayub, Q.~Zhao, M.~Meloy, E.~V. Sullivan, A.~Pfefferbaum, E.~Adeli, and K.~M.
  Pohl, ``Inpainting cropped diffusion mri using deep generative models,'' in
  \emph{International Workshop on PRedictive Intelligence In MEdicine}.\hskip
  1em plus 0.5em minus 0.4em\relax Springer, 2020, pp. 91--100.

\bibitem{tang2021lg}
Z.~Tang, M.~Cabezas, D.~Liu, M.~Barnett, W.~Cai, and C.~Wang, ``L{G}-{N}et:
  Lesion gate network for multiple sclerosis lesion inpainting,'' in
  \emph{International Conference on Medical Image Computing and
  Computer-Assisted Intervention (MICCAI)}.\hskip 1em plus 0.5em minus
  0.4em\relax Springer, 2021, pp. 660--669.

\bibitem{metz2017unrolled}
L.~Metz, B.~Poole, D.~Pfau, and J.~Sohl-Dickstein, ``Unrolled generative
  adversarial networks,'' in \emph{International Conference on Learning
  Representations (ICLR)}, 2017.

\bibitem{arjovsky2017wasserstein}
M.~Arjovsky, S.~Chintala, and L.~Bottou, ``Wasserstein generative adversarial
  networks,'' in \emph{International conference on machine learning
  (ICML)}.\hskip 1em plus 0.5em minus 0.4em\relax PMLR, 2017, pp. 214--223.

\bibitem{van2013wu}
D.~C. {Van Essen}, S.~M. {Smith}, D.~M. {Barch}, T.~E.~J. {Behrens},
  E.~{Yacoub}, and K.~{Ugurbil}, ``The {WU}-{M}inn human connectome project: an
  overview,'' \emph{Neuroimage}, vol.~80, pp. 62--79, 2013.

\bibitem{tustison2010n4itk}
N.~J. Tustison, B.~B. Avants, P.~A. Cook, Y.~Zheng, A.~Egan, P.~A. Yushkevich,
  and J.~C. Gee, ``{N4ITK}: improved {N3} bias correction,'' \emph{IEEE
  Transactions on Medical Imaging}, vol.~29, no.~6, pp. 1310--1320, 2010.

\bibitem{evans19933d}
A.~C. Evans, D.~L. Collins, S.~Mills, E.~D. Brown, R.~L. Kelly, and T.~M.
  Peters, ``3{D} statistical neuroanatomical models from 305 {MRI} volumes,''
  in \emph{1993 IEEE Conference Record Nuclear Science Symposium and Medical
  Imaging Conference}.\hskip 1em plus 0.5em minus 0.4em\relax IEEE, 1993, pp.
  1813--1817.

\bibitem{jenkinson2012fsl}
M.~Jenkinson, C.~F. Beckmann, T.~E. Behrens, M.~W. Woolrich, and S.~M. Smith,
  ``Fsl,'' \emph{Neuroimage}, vol.~62, no.~2, pp. 782--790, 2012.

\bibitem{andersson2015non}
J.~L. Andersson and S.~N. Sotiropoulos, ``Non-parametric representation and
  prediction of single-and multi-shell diffusion-weighted {MRI} data using
  gaussian processes,'' \emph{Neuroimage}, vol. 122, pp. 166--176, 2015.

\bibitem{andersson2016integrated}
J.~L. Andersson and S.~N. Sotiropoulos, ``An integrated approach to correction
  for off-resonance effects and subject movement in diffusion {MR} imaging,''
  \emph{Neuroimage}, vol. 125, pp. 1063--1078, 2016.

\bibitem{zeng2022fod}
R.~Zeng, J.~Lv, H.~Wang, L.~Zhou, M.~Barnett, F.~Calamante, and C.~Wang,
  ``{FOD-N}et: {A} deep learning method for fiber orientation distribution
  angular super resolution,'' \emph{Medical Image Analysis}, vol.~79, p.
  102431, 2022.

\bibitem{fischl2012freesurfer}
B.~Fischl, ``Free{S}urfer,'' \emph{Neuroimage}, vol.~62, no.~2, pp. 774--781,
  2012.

\bibitem{glasser2013minimal}
M.~F. Glasser, S.~N. Sotiropoulos, J.~A. Wilson, T.~S. Coalson, B.~Fischl,
  J.~L. Andersson, J.~Xu, S.~Jbabdi, M.~Webster, J.~R. Polimeni \emph{et~al.},
  ``The minimal preprocessing pipelines for the {H}uman {C}onnectome
  {P}roject,'' \emph{Neuroimage}, vol.~80, pp. 105--124, 2013.

\bibitem{behrens2003characterization}
T.~E. Behrens, M.~W. Woolrich, M.~Jenkinson, H.~Johansen-Berg, R.~G. Nunes,
  S.~Clare, P.~M. Matthews, J.~M. Brady, and S.~M. Smith, ``Characterization
  and propagation of uncertainty in diffusion-weighted {MR} imaging,''
  \emph{Magnetic Resonance in Medicine: An Official Journal of the
  International Society for Magnetic Resonance in Medicine}, vol.~50, no.~5,
  pp. 1077--1088, 2003.

\bibitem{kochunov2007relationship}
P.~Kochunov, P.~M. Thompson, J.~L. Lancaster, G.~Bartzokis, S.~Smith, T.~Coyle,
  D.~Royall, A.~Laird, and P.~T. Fox, ``Relationship between white matter
  fractional anisotropy and other indices of cerebral health in normal aging:
  tract-based spatial statistics study of aging,'' \emph{Neuroimage}, vol.~35,
  no.~2, pp. 478--487, 2007.

\bibitem{sexton2011meta}
C.~E. Sexton, U.~G. Kalu, N.~Filippini, C.~E. Mackay, and K.~P. Ebmeier, ``A
  meta-analysis of diffusion tensor imaging in mild cognitive impairment and
  {A}lzheimer's disease,'' \emph{Neurobiology of aging}, vol.~32, no.~12, pp.
  2322--e5, 2011.

\bibitem{cercignani2001mean}
M.~Cercignani, M.~Inglese, E.~Pagani, G.~Comi, and M.~Filippi, ``Mean
  diffusivity and fractional anisotropy histograms of patients with multiple
  sclerosis,'' \emph{American Journal of Neuroradiology}, vol.~22, no.~5, pp.
  952--958, 2001.

\bibitem{giulietti2018whole}
G.~Giulietti, M.~Torso, L.~Serra, B.~Span{\`o}, C.~Marra, C.~Caltagirone,
  M.~Cercignani, M.~Bozzali, and A.~D. N.~I. (ADNI), ``Whole brain white matter
  histogram analysis of diffusion tensor imaging data detects microstructural
  damage in mild cognitive impairment and alzheimer's disease patients,''
  \emph{Journal of Magnetic Resonance Imaging}, vol.~48, no.~3, pp. 767--779,
  2018.

\bibitem{rose2008gray}
S.~E. Rose, A.~L. Janke~Phd, and J.~B. Chalk, ``Gray and white matter changes
  in {A}lzheimer's disease: a diffusion tensor imaging study,'' \emph{Journal
  of Magnetic Resonance Imaging: An Official Journal of the International
  Society for Magnetic Resonance in Medicine}, vol.~27, no.~1, pp. 20--26,
  2008.

\bibitem{metwalli2010utility}
N.~S. Metwalli, M.~Benatar, G.~Nair, S.~Usher, X.~Hu, and J.~D. Carew,
  ``Utility of axial and radial diffusivity from diffusion tensor mri as
  markers of neurodegeneration in amyotrophic lateral sclerosis,'' \emph{Brain
  research}, vol. 1348, pp. 156--164, 2010.

\bibitem{melzer2013white}
T.~R. Melzer, R.~Watts, M.~R. MacAskill, T.~L. Pitcher, L.~Livingston, R.~J.
  Keenan, J.~C. Dalrymple-Alford, and T.~J. Anderson, ``White matter
  microstructure deteriorates across cognitive stages in {P}arkinson disease,''
  \emph{Neurology}, vol.~80, no.~20, pp. 1841--1849, 2013.

\bibitem{schmierer2007diffusion}
K.~Schmierer, C.~A. Wheeler-Kingshott, P.~A. Boulby, F.~Scaravilli, D.~R.
  Altmann, G.~J. Barker, P.~S. Tofts, and D.~H. Miller, ``Diffusion tensor
  imaging of post mortem multiple sclerosis brain,'' \emph{Neuroimage},
  vol.~35, no.~2, pp. 467--477, 2007.

\end{thebibliography}

\end{document}